\documentclass[pdftex,twocolumn,nofootinbib,preprintnumbers,superscriptaddress]{revtex4-1}

\RequirePackage[T1]{fontenc}

\RequirePackage{graphicx}
\RequirePackage{mathptmx}      
\RequirePackage{flushend}
\usepackage{amsmath}
\usepackage{amssymb}
\usepackage{float}
\usepackage{caption,subcaption}
\usepackage{multirow}
\usepackage{comment}
\begin{document}

\title{Measurement of the helicity dependence for single $\pi^{0}$ photoproduction from the deuteron}



\author{F.~Cividini}
\affiliation{Institut f\"ur Kernphysik, University of Mainz, D-55099 Mainz, Germany}\label{Mainz}

\author{M.~Dieterle}
\affiliation{Departement Physik, Universit\"at Basel, CH-4056 Basel, Switzerland}
\author{ S.~Abt}
\affiliation{Departement Physik, Universit\"at Basel, CH-4056 Basel, Switzerland}

\author{ P.~Achenbach}
\affiliation{Institut f\"ur Kernphysik, University of Mainz, D-55099 Mainz, Germany}\label{Mainz}

\author{P.~Adlarson}
\affiliation{Institut f\"ur Kernphysik, University of Mainz, D-55099 Mainz, Germany}

\author{F.~Afzal}
\affiliation{Helmholtz-Institut f\"ur Strahlen- und Kernphysik, Universit\"at Bonn, D-53115 Bonn, Germany}

\author{Z.~Ahmed}
\affiliation{University of Regina, Regina, Saskatchewan S4S 0A2, Canada}

\author{J.R.M.~Annand}
\affiliation{SUPA School of Physics and Astronomy, University of Glasgow, Glasgow G12 8QQ, United Kingdom}

\author{H.J.~Arends}
\affiliation{Institut f\"ur Kernphysik, University of Mainz, D-55099 Mainz, Germany}

\author{M.~Bashkanov}
\affiliation{Department of Physics, University of York, Heslington, York, Y010 5DD, UK}

\author{R.~Beck}
\affiliation{Helmholtz-Institut f\"ur Strahlen- und Kernphysik, Universit\"at Bonn, D-53115 Bonn, Germany}

      \author{M.~Biroth}\affiliation{Institut f\"ur Kernphysik, University of Mainz, D-55099 Mainz, Germany}
      
      \author{N.~Borisov}
\affiliation{Joint Institute for Nuclear Research, 141980 Dubna, Russia}

      \author{A.~Braghieri}\affiliation{INFN Sezione di Pavia, I-27100 Pavia, Italy}%
      \author{W.J.~Briscoe}\affiliation{The George Washington University, Washington, DC 20052-0001, USA}

      \author{C.~Collicott}\affiliation{Institut f\"ur Kernphysik, University of Mainz, D-55099 Mainz, Germany}
      
      \author{S.~Costanza}
\affiliation{INFN Sezione di Pavia, I-27100 Pavia, Italy}%
\affiliation{Dipartimento di Fisica, Universit\`a di Pavia, Pavia; Italy}

\author{A.~Denig}\affiliation{Institut f\"ur Kernphysik, University of Mainz, D-55099 Mainz, Germany}

      \author{A.S.~Dolzhikov}
      \affiliation{Joint Institute for Nuclear Research, 141980 Dubna, Russia}
     
      \author{E.J.~Downie}\affiliation{The George Washington University, Washington, DC 20052-0001, USA}
      
         \author{P.~Drexler}\affiliation{Institut f\"ur Kernphysik, University of Mainz, D-55099 Mainz, Germany}

\author{S.~Fegan}
\affiliation{Department of Physics, University of York, Heslington, York, Y010 5DD, UK}

      \author{A.~Fix}\affiliation{Tomsk Polytechnic University, 634034 
Tomsk, Russia}

      \author{S.~Gardner}\affiliation{SUPA School of Physics and Astronomy, University of Glasgow, Glasgow G12 8QQ, United Kingdom}
      
      \author{D.~Ghosal}\affiliation{Departement Physik, Universit\"at Basel, CH-4056 Basel, Switzerland}
      
      \author{D.I.~Glazier}\affiliation{SUPA School of Physics and Astronomy, University of Glasgow, Glasgow G12 8QQ, United Kingdom}

      \author{I.~Gorodnov}\affiliation{Joint Institute for Nuclear Research, 141980 Dubna, Russia}
      
      \author{W.~Gradl}
\affiliation{Institut f\"ur Kernphysik, University of Mainz, D-55099 Mainz, Germany}

      \author{M.~G\"unther}\affiliation{Departement Physik, Universit\"at Basel, CH-4056 Basel, Switzerland}
      
      \author{D.~Gurevich}\affiliation{Institute for Nuclear Research, 125047 Moscow, Russia}
      
      \author{L.~Heijkenskj\"old}\affiliation{Institut f\"ur Kernphysik, University of Mainz, D-55099 Mainz, Germany}
      
      \author{D.~Hornidge}\affiliation{Mount Allison University, Sackville, New Brunswick E4L 1E6, Canada}
      
      \author{G.M.~Huber}\affiliation{University of Regina, Regina, Saskatchewan S4S 0A2, Canada}
      
      \author{A.~K\"aser}\affiliation{Departement Physik, Universit\"at Basel, CH-4056 Basel, Switzerland}
      
      \author{V.L.~Kashevarov}
       \affiliation{Institut f\"ur Kernphysik, University of Mainz, D-55099 Mainz, Germany}
            \author{S.J.D.~Kay}\affiliation{University of Regina, Regina, Saskatchewan S4S 0A2, Canada}
      
      \author{M.~Korolija}\affiliation{Rudjer Boskovic Institute, HR-10000 Zagreb, Croatia}
      
      \author{B.~Krusche}\affiliation{Departement Physik, Universit\"at Basel, CH-4056 Basel, Switzerland}
      
      \author{A.~Lazarev}\affiliation{Joint Institute for Nuclear Research, 141980 Dubna, Russia}
      \author{K.~Livingston}\affiliation{SUPA School of Physics and Astronomy, University of Glasgow, Glasgow G12 8QQ, United Kingdom}
      \author{S.~Lutterer}\affiliation{Departement Physik, Universit\"at Basel, CH-4056 Basel, Switzerland}
      \author{I.J.D.~MacGregor}\affiliation{SUPA School of Physics and Astronomy, University of Glasgow, Glasgow G12 8QQ, United Kingdom}
      
      \author{D.M.~Manley}\affiliation{Kent State University, Kent, Ohio 44242-0001, USA}
      
      \author{P.P.~Martel}\affiliation{Institut f\"ur Kernphysik, University of Mainz, D-55099 Mainz, Germany}
      
      \author{R.~Miskimen}\affiliation{University of Massachusetts, Amherst, Massachusetts 01003, USA}
      
      \author{M.~Mocanu}\affiliation{Department of Physics, University of York, Heslington, York, Y010 5DD, UK}
      
      \author{E.~Mornacchi}\affiliation{Institut f\"ur Kernphysik, University of Mainz, D-55099 Mainz, Germany}
      
      \author{C.~Mullen}\affiliation{SUPA School of Physics and Astronomy, University of Glasgow, Glasgow G12 8QQ, United Kingdom}
      
      \author{A.~Neganov}\affiliation{Joint Institute for Nuclear Research, 141980 Dubna, Russia}
      
      \author{A.~Neiser}\affiliation{Institut f\"ur Kernphysik, University of Mainz, D-55099 Mainz, Germany}
      \author{M.~Ostrick}\affiliation{Institut f\"ur Kernphysik, University of Mainz, D-55099 Mainz, Germany}

      \author{D.~Paudyal}\affiliation{University of Regina, Regina, Saskatchewan S4S 0A2, Canada}
      
      \author{P.~Pedroni}\email{E-mail:paolo.pedroni@pv.infn.it}
      \affiliation{INFN Sezione di Pavia, I-27100 Pavia, Italy}
      
      \author{A.~Powell}\affiliation{SUPA School of Physics and Astronomy, University of Glasgow, Glasgow G12 8QQ, United Kingdom}
      
      \author{T.~Rostomyan}
      \altaffiliation{Now at Paul Scherrer Institute (PSI), CH-5232 Villigen PSI, Switzerland.}
\affiliation{Departement Physik, Universit\"at Basel, CH-4056 Basel, Switzerland}
      
       \author{V.~Sokhoyan}\affiliation{Institut f\"ur Kernphysik, University of Mainz, D-55099 Mainz, Germany}
       
      \author{K.~Spieker}\affiliation{Helmholtz-Institut f\"ur Strahlen- und Kernphysik, Universit\"at Bonn, D-53115 Bonn, Germany}
      
      \author{O.~Steffen}\affiliation{Institut f\"ur Kernphysik, University of Mainz, D-55099 Mainz, Germany}

\author{I.~Strakovsky}\affiliation{The George Washington University, Washington, DC 20052-0001, USA}

      \author{T.~Strub}\affiliation{Departement Physik, Universit\"at Basel, CH-4056 Basel, Switzerland}
      \author{A.~Thiel}\affiliation{Helmholtz-Institut f\"ur Strahlen- und Kernphysik, Universit\"at Bonn, D-53115 Bonn, Germany}
      
      \author{M.~Thiel}\affiliation{Institut f\"ur Kernphysik, University of Mainz, D-55099 Mainz, Germany}
      \author{A.~Thomas}\affiliation{Institut f\"ur Kernphysik, University of Mainz, D-55099 Mainz, Germany}
      
      \author{Yu.A.~Usov}\affiliation{Joint Institute for Nuclear Research, 141980 Dubna, Russia}
      
      \author{S.~Wagner}\affiliation{Institut f\"ur Kernphysik, University of Mainz, D-55099 Mainz, Germany}
      \author{D.P.~Watts}\affiliation{Department of Physics, University of York, Heslington, York, Y010 5DD, UK}
      
      \author{D.~Werthm\"uller}
\altaffiliation{Now at Paul Scherrer Institute (PSI), CH-5232 Villigen PSI, Switzerland}
\affiliation{Department of Physics, University of York, Heslington, York, Y010 5DD, UK}

      \author{J.~Wettig} \affiliation{Institut f\"ur Kernphysik, University of Mainz, D-55099 Mainz, Germany}
      
      \author{L.~Witthauer}\affiliation{Departement Physik, Universit\"at Basel, CH-4056 Basel, Switzerland}
      
      \author{M.~Wolfes}\affiliation{Institut f\"ur Kernphysik, University of Mainz, D-55099 Mainz, Germany}
      \author{N.~Zachariou}\affiliation{Department of Physics, University of York, Heslington, York, Y010 5DD, UK}

\collaboration{The A2 Collaboration at MAMI}

\begin{abstract}

  The helicity-dependent single $\pi^{0}$ photoproduction cross section on the deuteron and the angular dependence of the double polarisation observable $E$ for the quasi-free single $\pi^0$ production off the proton and the neutron have been measured, for the first time, from the threshold region up to the photon energy 1.4 GeV.
  The experiment was performed at the tagged photon facility of the MAMI accelerator and used a circularly polarised photon beam and longitudinally polarised deuteron target. The reaction products were detected using the large acceptance Crystal Ball/TAPS calorimeter, which covered 97\% of the full solid angle.
  Comparing the cross section from the deuteron with the sum of free nucleon
  cross sections provides a quantitative estimate of the effects of the
  nuclear medium on pion production.
  In contrast, comparison
  of the $E$ helicity asymmetry data from quasi-free protons off deuterium
  with data from a
  free proton target indicates that nuclear effects do not significantly
  affect this observable.
  As a consequence, it is deduced that the helicity asymmetry $E$ on a free neutron can be reliably extracted from measurements on a deuteron in quasi-free kinematics. 

\end{abstract}

\maketitle


\section{Introduction}

Despite many decades spent in intense research,
many open questions remain regarding the the structure of the nucleon.
The strong interaction plays a decisive role in the internal dynamics of the nucleon and its excited states, similar to the way the electromagnetic interaction relates to the fundamental properties of atomic excitation spectra. Therefore, the study of the nucleon's spectrum is a crucial step towards understanding its structure.

The resonance widths are determined by the strong interaction and are of the order of hundreds of MeV, whereas their spacing is no more than a few tens of MeV, which leads to a very large amount of overlapping.
To disentangle and access the individual states in the nucleon's spectrum, measurements of different polarisation observables are of crucial importance.

In general, single meson photoproduction on a nucleon can be described
using 8 spin amplitudes, dependent
 on the photon-nucleon helicity configurations and on
the total center-of-mass energy.
Due to parity conservation, the number of independent amplitudes reduces to 4. Since each observable has a Hermitian form in the amplitudes, there are 16 linearly independent observables. They can be accessed using different combinations of polarisation of the photon beam, the target and recoil nucleon polarisation as, for instance, discussed in detail in Ref.~\cite{donna}.

According to different theoretical studies (see, for example, Refs.~\cite{keat,chang,beck,thiel}),
 it is sufficient to measure a limited set (no less than 8) of properly chosen observables
to unambiguously determine all four spin amplitudes for single pion photoproduction.
Furthermore, since the electromagnetic interaction does not conserve isospin, it is necessary to use both proton and neutron targets
in order to access the isospin decomposition of the amplitudes.
In light of these points, the A2 Collaboration has performed a series of experiments in order to measure a range of different
polarisation observables at photon energies up 1500 MeV, both on the proton and on the neutron.

In the present paper, we extract, for the first time, the angular dependence
of the helicity asymmetry $E$ for single $\pi^0$ production on the neutron.
Data
were collected using a polarised photon beam along with a polarised nucleon target.
When the polarisation of the recoil nucleons is not measured,
the corresponding polarised cross section may be written:
%
%
\begin{eqnarray}\label{target-beam}
  &&\frac{d\sigma}{d\Omega}= \frac{d\sigma_0}{
    d\Omega}\Big\{1-P_{L}^{\gamma}\Sigma\cos(2\phi)\nonumber\\
&&\phantom{xxxxxx}+P_{x}^{T}\big[-P_{L}^{\gamma}H\sin(2\phi)+P_{\odot}^{\gamma}F \big]\\
&&\phantom{xxxxxxxx}+P_{y}^{T}\bigg(T-P_{L}^{\gamma}P\cos(2\phi)\bigg)\nonumber\\
&&\phantom{xxxxxxxxxx}+P_{z}^{T}\big[P_{L}^{\gamma}G\sin(2\phi)-P_{\odot}^{\gamma}E\big]\Big\},\nonumber
\end{eqnarray}
where $\sigma_0$ is the unpolarised cross section.
The notation $P_{L}^{\gamma}$ ($P_{\odot}^{\gamma}$) refers to the linear (circular) polarisation of the photon beam, 
while $P_i^{T}$, $i=x,y,z$, stands for degree of the target polarisation,  with $P_{z}^{T}$ being
its longitudinal polarisation.
The observables ${\cal O}= \Sigma,H,F,\ldots$ are the standard polarisation asymmetries and
$\phi$ is the angle between the linear photon polarisation plane and the reaction plane. The latter is defined by the incident photon momentum and the momentum of the outgoing pion.

During data taking, the photon beam was circularly polarised and the target was longitudinally polarised.
With these experimental conditions, Eq.~(\ref{target-beam}) reduces to:
%
%
\begin{equation}
\frac{d\sigma^{\uparrow\downarrow/\uparrow\uparrow}}{d\Omega}= \frac{d\sigma_0}{d\Omega}\,\Big\{1 \pm P_{z}^{T}P_{\odot}^{\gamma}E\Big\},
\label{E_obs1}
\end{equation}
where the notation $\uparrow\uparrow(\uparrow\downarrow)$
indicates the relative parallel (anti-parallel) photon-target polarisation direction.
From Eq.\,(\ref{E_obs1}) the double polarisation observable $E$ can also be presented as:
\begin{eqnarray}\label{E_obs2}
  &&E=\frac{d\sigma^{\uparrow\downarrow}/d\Omega-d\sigma^{\uparrow\uparrow}/d\Omega}
  {d\sigma^{\uparrow\downarrow}/d\Omega+d\sigma^{\uparrow\uparrow}/d\Omega} = \nonumber\\
 &&\phantom{x}=
  \frac{N^{\uparrow\downarrow}-N^{\uparrow\uparrow}}{N^{\uparrow\downarrow}+N^{\uparrow\uparrow}}
  \cdot\frac{1}{P_{z}^{T}}\cdot\frac{1}{P_{\odot}^{\gamma}}\cdot\frac{1}{d}\ ,
\end{eqnarray}
where 
%
%
$N^{^{\uparrow\downarrow (\uparrow\uparrow)}}$ indicates the number of events with a parallel
(anti-parallel) photon-target helicity configuration,
and $d$ is the dilution factor describing the fraction
of polarised nucleons inside the target.

In view of the impossibility of creating an appropriate free neutron target, one has to rely on light nuclei, such as $^3$He or deuterium, as effective neutron targets.
    This choice minimizes the nuclear corrections
    both due to off-shell neutron effects and to final-state interactions (FSI).

For the present experiment, a deuterated butanol (C$_4$D$_9$OD) target was chosen, which can reach     
high degrees of polarisation (up to about $70\%$)
with a fast build-up time ($\sim$~some hours)
and high relaxation times (several hundred hours). 
Compared to other target materials like ammonia, butanol has the advantage that
the background C and O nuclei, being spinless, are not polarised.

For an unambiguous extraction of the cross section on a single nucleon, reliable control of various nuclear effects is required. In order to minimize their influence one uses, as a rule, quasi-free kinematics,
where the incoming photon interacts with only a single nucleon 
and the other nucleons
    may be regarded as spectators.

In addition, a robust theoretical model that takes into account the most important nuclear effects such as Fermi motion, admixture of tensor forces, the Pauli exclusion principle and FSI, is also of vital importance.
    For a reliable control and modeling of nuclear effects,
    different additional $N\pi(\pi)$ reaction channels on different nuclear targets
    need also to be measured in order to obtain a comprehensive  experimental quantitative evaluation of their impact on the measured data.

In recent years, some of these issues have already been addressed in 
theoretical and experimental studies of other
observables. For instance, 
the GWU-ITEP theoretical group has shown~\cite{review,itep1,itep2,itep3}
that FSI corrections decrease
the unpolarised cross section of the 
$N\pi$ reactions on the deuteron, compared to the free nucleon case,
by up to 20\%,
while  their effect is much smaller and consistent
with experimental uncertainties for the different polarisation asymmetries observables.

In particular, the model described in Ref.~\cite{itep2}
was used to extract
 the unpolarised differential cross section for the $\gamma n \to n\pi^0$ 
 reaction from our previous measurement on a deuteron target~\cite{SMA19}
 in the photon energy range $200\ \!-\ \! 813$~MeV.
 Above 300~MeV,  a satisfactory agreement was found between the extracted data
 and the SAID-MA19 partial wave analysis.

 On the other hand, our recent data for the photon beam asymmetry
 of the $\pi^0$ production off neutrons bound in deuterons from 390 to
 610~MeV~\cite{mullen}
 are well reproduced by existing partial wave analyses for the lowest
 measured region while some discrepancies appear only at the highest energies.
 This feature seems to indicate that nuclear effects have a
 smaller effect on polarization asymmetries since these observables measure
 ratios of absolute cross sections.

    It is interesting to note that
    similar conclusions have been drawn from our previous studies
    on the beam-helicity asymmetry of both $\eta\pi$ and
    $\pi^0 \pi^\pm$ pairs off protons and 
    deuterons~\cite{kaser,ober}.  In these cases,
    measured unpolarised absolute cross sections are decreased for the
    reactions on quasi-free protons on the deuteron
    with respect to the free proton data while the measured asymmetries
    do not show significant differences between these two cases.
    Moreover, the measured beam-helicity asymmetry
    of $\eta\pi^0$ pair on C, Al and Pb~\cite{sokh} has been found to be 
    not affected significantly by nuclear effects. 

    The main goal of the present work, which extends these studies to
    helicity-dependent observables, is  twofold:
(i) to measure the helicity-dependent semi-inclusive cross section for single $\pi^0$ production on the deuteron to clearly single out the role of nuclear effects and (ii) to measure the $E$ observable for single $\pi^0$ production on quasi-free neutrons and protons. In the latter case, the comparison with the data on a free proton provides a cross check to elucidate  the influence of the nuclear environment on the single nucleon process. These new results extend the set of the angle-integrated $E$ data that have already been published
in Ref.~\cite{diet1}.

\section{Experimental setup}\label{expapp}

\begin{figure*}%
\centering
\includegraphics[scale=0.27]{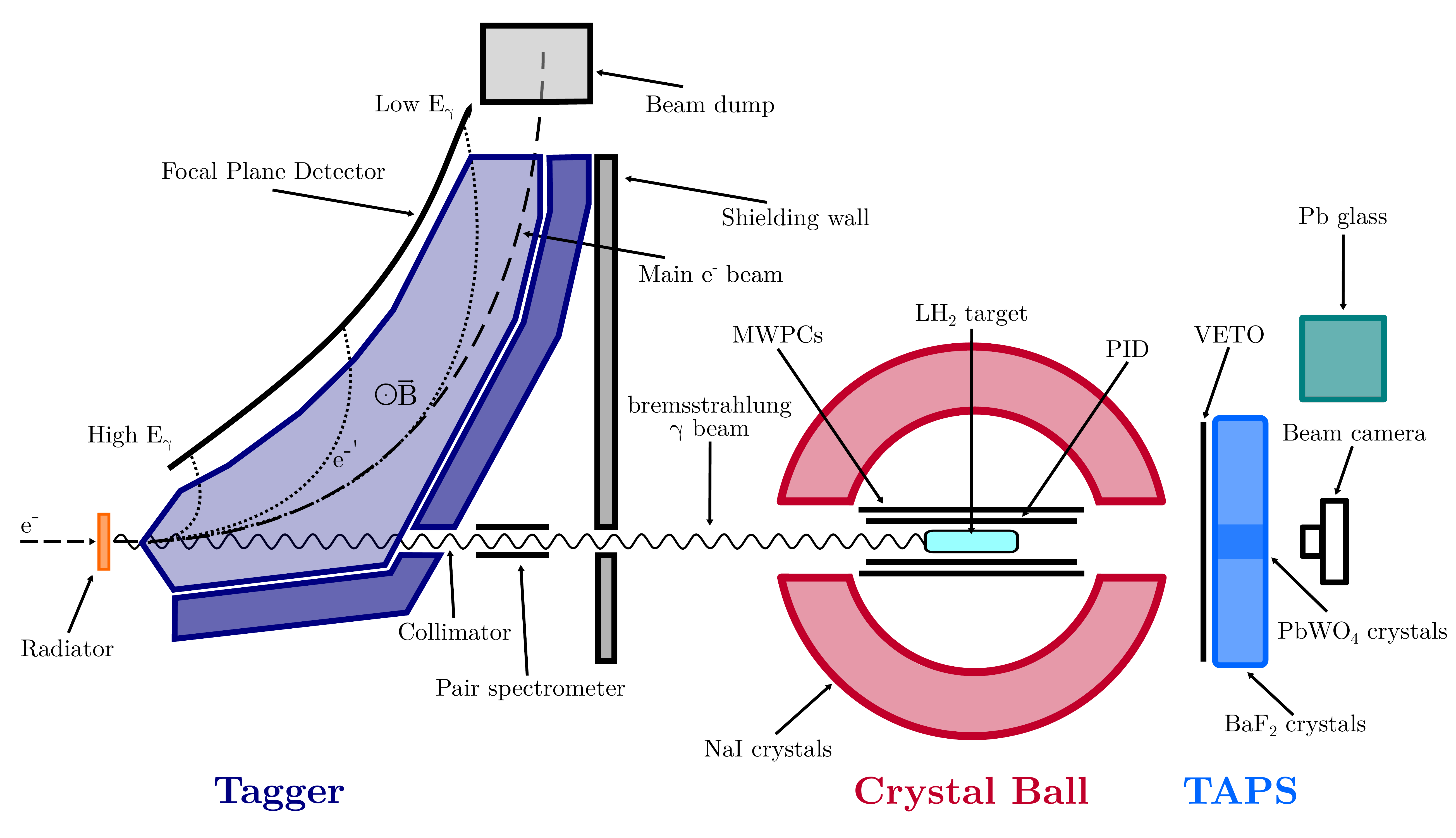}
\caption{A2 experimental setup with the photon tagging apparatus and
  detectors~\cite{edo}. }
\label{fig_A2_setup}
\end{figure*}

%


The helicity-dependent data were measured during different beam time periods at the MAMI electron
accelerator facility in Mainz, Germany~\cite{mamic}.
Figure~\ref{fig_A2_setup} shows a general sketch of the A2 experimental setup used for the measurement. Since this apparatus has been already previously described in detail (see, for instance, Refs.~\cite{A2:2013wkp,A2:2014pie,diet1,witt1,diet2} and references therein), only the main characteristics relevant for the present measurement will be given here.

The photon beam was produced via bremsstrahlung when the primary polarised electron beam hits a thin amorphous radiator.
To avoid polarisation-dependent photon flux values,
the beam helicity was  flip\-ped at a rate of 1~Hz.

The electron polarisation $P_e$ was regularly  
determined by Mott scattering close to the electron source and found to 
be around 80\% for all the different measurements.
In addition, Moeller scattering at the radiator site was used as an additional polarisation monitor.
A magnetic field applied after the radiator deflected the post-brems\-strahlung electrons in the focal plane. Electrons were
tagged by the Glasgow-Mainz spectrometer with an energy resolution of $\sim2-5$~MeV, which corresponds to the width of the
focal plane counters~\cite{taggnew}.
The photon beam passed through a 2~mm diameter collimator, reaching the target and detection  apparatus.

The degree of energy-dependent circular photon polarisation, $P_{\odot}^{\gamma}$
was determined using the Olsen and Maximon formula~\cite{olsen}:
\begin{equation*}
  \frac{P_{\odot}^{\gamma}}{p_e} = \frac{4x-x^2}{4-4x+3x^2} \ ,   
\end{equation*}
where $x=  E_e / E_\gamma$, and $E_e$, $E_\gamma$  are the energy of the electron and of the
bremsstrahlung photon, respectively.     
The polarisation degree was highest for maximum photon energies and diminished with decreasing energies. 
 
The photon tagging efficiency (approximately $35\%$) was measured once a day using a Pb-Glass Cerenkov detector in dedicated low flux runs. During the standard data
taking operation, the fluctuations of the photon flux were monitored using a low-efficiency pair spectrometer
located in the photon beamline after the collimator.
An absolute systematic uncertainty in the photon flux of 4\% has been
estimated from the comparison of the data from these detectors obtained under
a range of different experimental conditions.

The target used for this experiment was the Mainz-Dubna Frozen Spin Target (FST) filled with deuterated butanol~\cite{Rohl,Thomas}. 
The filling factor for the $\sim 2$ mm diameter butanol
spheres included
in the 2 cm long, 2 cm diameter target container was estimated to be  60\%, with a systematic uncertainty of 2\%~\cite{Rohl}.
The target material was polarised using the Dynamic Nuclear Polarisation (DNP) effect~\cite{Brad99}, which required a magnetic field
 $B = 1.5$~T and a temperature of  $\sim$25 mK.
Such conditions, in combination with a small holding magnetic field of 0.6~T which replaced the polarising magnet during the data taking phase, allowed regular relaxation times longer than $>$1000~h
to be obtained.
The target polarisation was measured with an NMR system before and after the data taking period and then interpolated exponentially at intermediate
times.

    To enhance the efficiency of the DNP procedure, the butanol was chemically doped with highly polarisable paramagnetic centers. In the first two beam times,
 the trityl radical Finland D36 was used, with 
typical polarisation degrees of about 60\%.
However, for these runs there was a problem in the absolute determination of the polarisation,
caused by small field inhomogeneities ($\Delta B \leq 1.78$~mT) of the polarising magnet.

    To solve this problem,
    an additional beam time
    used a different  radical (Tem\-po), which resulted in lower polarisation degrees (about 30\%), but was not sensitive to small field
inhomogeneities. Therefore,
the absolute scale of all helicity-dependent cross sections and asymmetries
obtained in the previous runs were renormalized to this final beam time,
which comprised about 30\% of the total collected statistics.
    The evaluated correction factors were also
    cross checked with a parallel analysis on $\eta$ photoproduction at threshold since,
        in this case, the $E$ asymmetry equals one, due to the predominant
     $s$-wave production
     mechanism~\cite{A2:2016bij,diet1}.
From these analyses, similar to those in Ref.~\cite{diet1}, a conservative relative systematic uncertainty of 10\% has been estimated for the degree of target polarisation.

For the evaluation of the denominator of Eq.~(\ref{E_obs2}), it was also
crucial to study the contribution of the
unpolarised C and O  nuclei inside the target material.
Some dedicated data runs with a carbon target were performed for this purpose. This target was made from foam with the same density and the same geometry as the butanol target.

Photons from $\pi^0$ decay and recoil nucleons were detected by
the Crystal Ball-TAPS apparatus. The Crystal Ball (CB) wass located around the target cell and covered the full azimuthal ($\phi$) angle and  polar ($\theta$) from 21$^{\circ}$ to 159$^{\circ}$~\cite{artcb}.
It consisted of 672 NaI(Tl) crystals covering a large solid angle and
was $\simeq 100\%$ detection efficient for photons coming from the $\pi^0$ decay.
Inside the CB there were two Multi-Wire Proportional Chambers (MWPCs) and a Particle Identification Detector (PID), made of a barrel of 24 plastic scintillators. The combination of all these detectors provided a precise tracking and identification of charged particles. 
TAPS was a hexagonal wall covering the polar forward region outside the CB acceptance and was made of 366 BaF$_2$ and 72 PbWO$_{4}$ crystals~\cite{taps1,taps2}.
In front of the TAPS array a 5 mm thick plastic scintillator wall (VETO) was used for charged particles identification.
The combination of the large acceptance CB and TAPS covered $\sim$ 97$\%$ of the full solid angle. 

    Two different experimental triggers  were used to collect the data presented
    here.
    A general-purpose trigger
        required the total sum of pulse amplitudes  from the CB or TAPS crystals
        to exceed a hardware threshold corresponding to $\simeq 50$~MeV~\cite{fede}.
        A second trigger, optimized for the selection of single $\pi^0$
        events at photon energies above $\simeq 450$~MeV,
        required a higher energy threshold ( $\simeq 250$~MeV),
        with  the additional conditions to have
        at least two  hardware clusters (groups of adjacent hit crystals)
        in CB and TAPS together~\cite{manu,diet1}.
    
\section{Data Analysis}

After the energy and time calibration of all detector modules,
the data from the butanol target were analysed together with data from the carbon foam.
All  the different algorithms used to analyse the collected data
have been tested and checked  with simulation, to obtain an optimal rejection of the background coming from the unpolarised target nucleons.

Detailed descriptions of these algorithms have been given before
(see, for instance Refs.~\cite{diet1,witt1,diet2} and references therein).
Therefore, only a summary of the main analysis steps needed for the identification of the measured observables on the $\pi^0$ production
on deuterium will be given here.


The $\pi^{0}$ identification algorithm was common to all
the offline analyses of the collected data, while the methods for
 nucleon identification and for the subtraction of the unpolarised background were  only used for the evaluation of the $E$ observable.

    The detector response and the efficiency of  reconstruction of
    single $\pi^0$ events, needed for the
 determination of the absolute cross section, were
 evaluated using a GEANT4 based simulation~\cite{geant4} which modelled
 accurately
the geometry and composition of the detector setup and
accounted for electronic thresholds.

    The candidate events accepted for the evaluation of both the differential cross section and the $E$ asymmetry
    were those with 2 or 3 
    clusters of energy deposition reconstructed inside of the detection apparatus.

\subsection{$\pi^{0}$  reconstruction and identification}

The first offline analysis step was the evaluation of the two photon  invariant mass (IM) using all the neutral clusters of each event. For all events with more than 2 neutral hits, all possible combinations were used to calculate the two photon invariant mass and only the combination giving the closest value to the nominal  $\pi^0$ mass was retained for successive analysis steps.

\begin{figure}%
\centering
\includegraphics[scale=0.38]{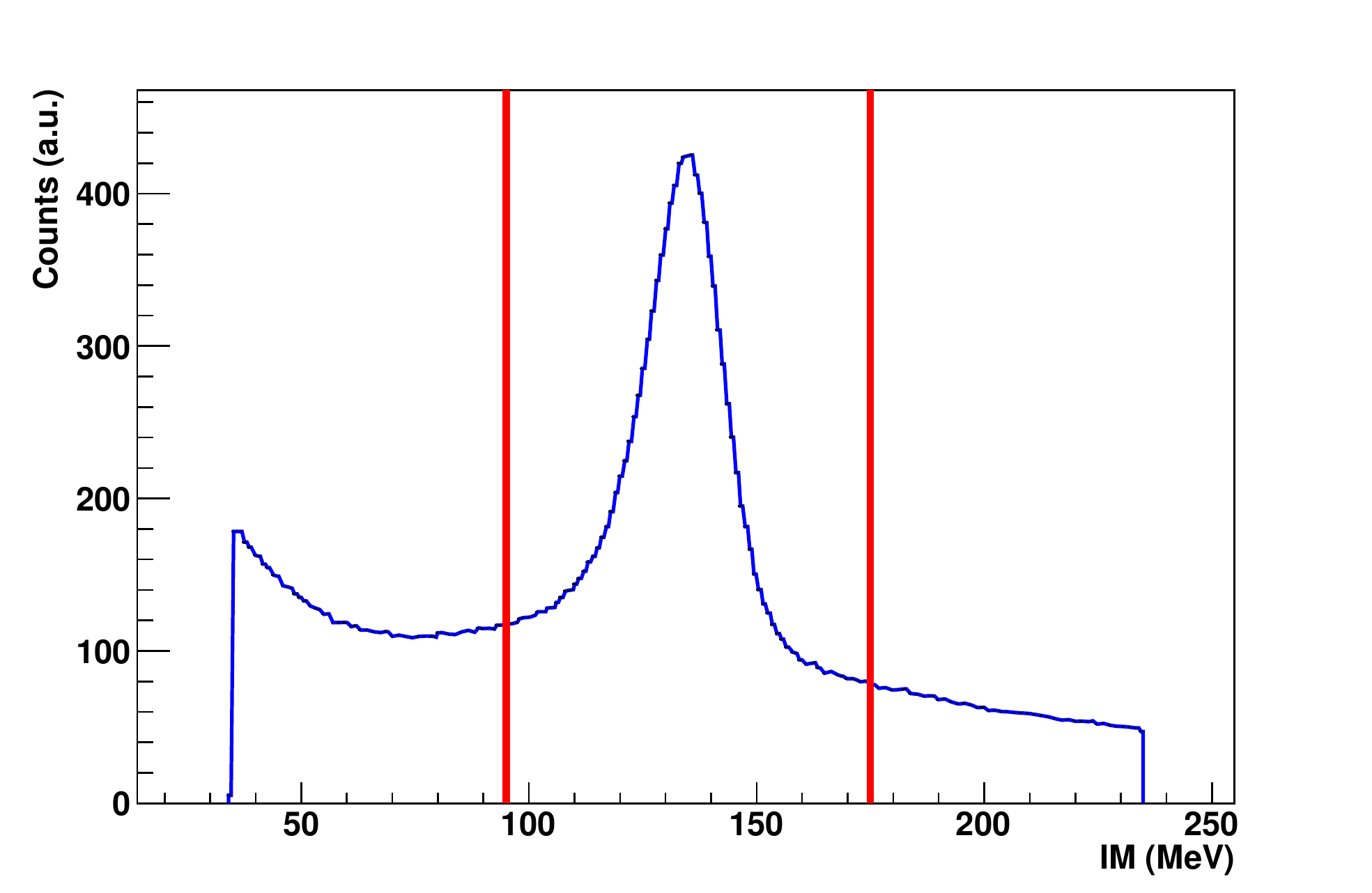}
\caption{Two photon invariant mass (IM) distribution from the $\pi^{0}$ reconstruction procedure  for all candidate events. The red lines define the $\pm$ 40 MeV cut (corresponding to an experimental resolution of about $2.5\ \sigma$) around the nominal $m_{\pi^0}$ value.}
\label{figure_IM}
\end{figure}

The event was selected for the next step
if the calculated IM  value was within $\pm 40$ MeV of the PDG $\pi^{0}$ nominal mass. This corresponds to an experimental resolution of about $2.5\ \sigma$.
In Fig.~\ref{figure_IM}, the overall IM distribution is shown
together with the applied invariant mass cut.

    For events with more than two neutral hits, where  
    ambiguities between photons and neutrons can occur,
    an additional test was performed by comparing the invariant mass of the
    two photon candidate $m_{\gamma_1\gamma_2}$ to the nominal $\pi^0$ mass $m_{\pi^0}$
    as:

    \begin{equation} \label{eqchi2}
    \chi^2_{\pi^0} = \left( 
    \frac{m_{\gamma_1\gamma_2}-m_{\pi^0}}{\Delta m_{\gamma_1\gamma_2}} \right)^2 \ ,
\end{equation}
     where $\Delta m_{\gamma_1\gamma_2}$ represents the uncertainty
     on $ m_{\gamma_1\gamma_2}$ due to the
    experimental resolution, as determined 
    by the simulated detector response.

    The two neutral clusters from 
the combination with the lowest $\chi^2_{\pi^0}$ value were selected as
$\pi^0$ decay photons and the remaining neutral hit as a neutron candidate.
As previously shown (see Refs.~\cite{diet1,diet3} and references therein), 
this method has proven to be very effective in resolving ambiguities
in neutron-photon separation both in the $n\pi^0$ and in the $n\pi^0\pi^0$
final states.

Only events with a reconstructed $\pi^0$ and with an IM value within the
selected window were accepted for the subsequent analysis steps. 

The following step
was the evaluation of the event  missing mass (MM), where the recoil nucleon of the 
reaction $\gamma  N \rightarrow \pi^{0} N$ was considered as a missing particle, even when it had been detected.
This parameter was calculated  as follows:
\begin{equation}
{\rm MM}=\sqrt{(E_{\gamma}+m_{N}-E_{\pi^{0}})^{2}-(\overrightarrow{p}_{\gamma}-\overrightarrow{p}_{\pi^{0}})^{2}} \ \ , 
\label{MM_equation}
\end{equation}
where $E_\gamma$ and $p_\gamma$ are the laboratory energy and momentum of the incoming photon, $m_N$ is the nucleon mass in the initial state,   $E_{\pi^0}$ and  $p_{\pi^0}$  are the reconstructed $\pi^{0}$ total energy and momentum.

\begin{figure}[ht]
\centering
\includegraphics[scale=0.35]{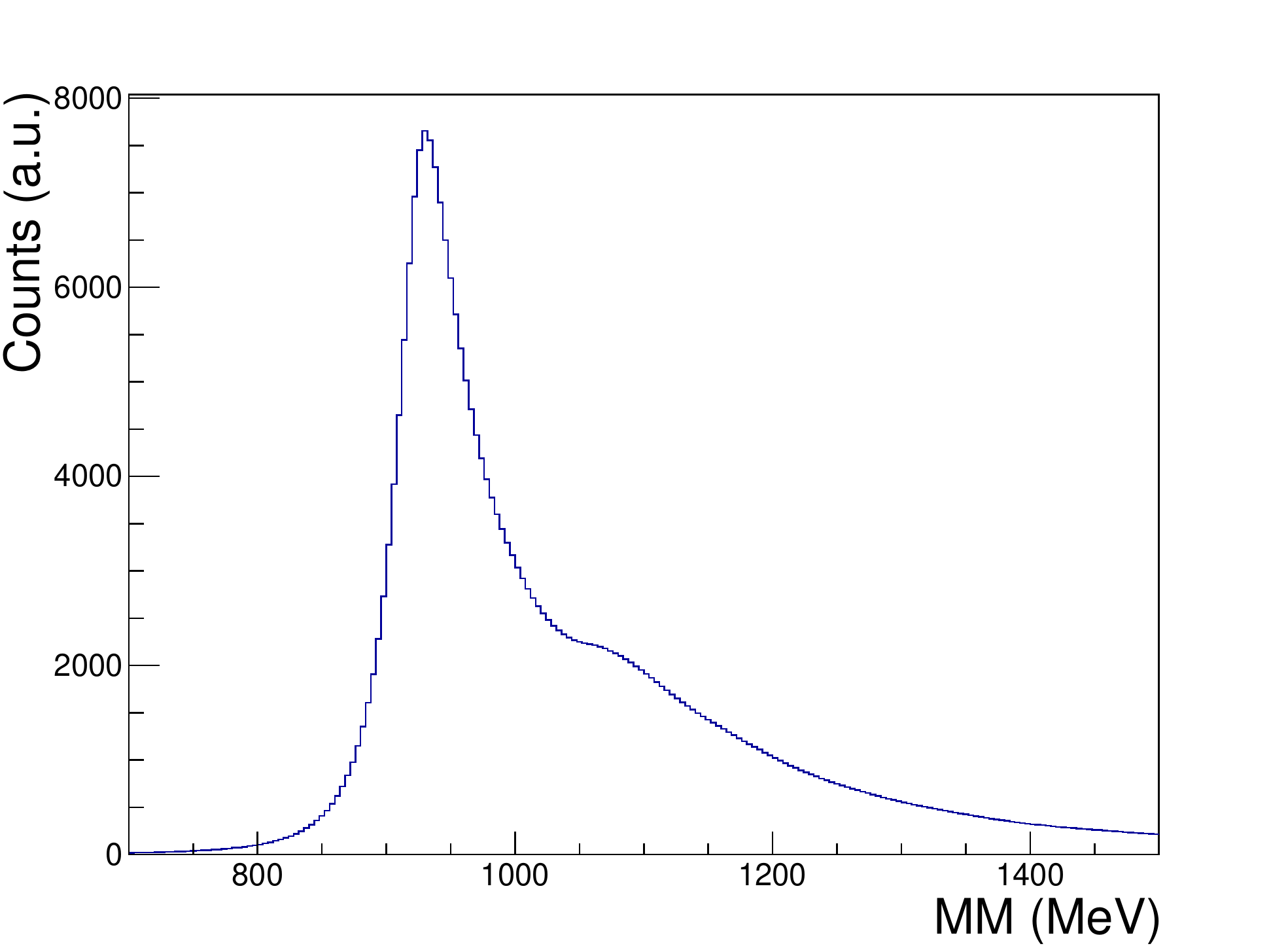}
\includegraphics[scale=0.35]{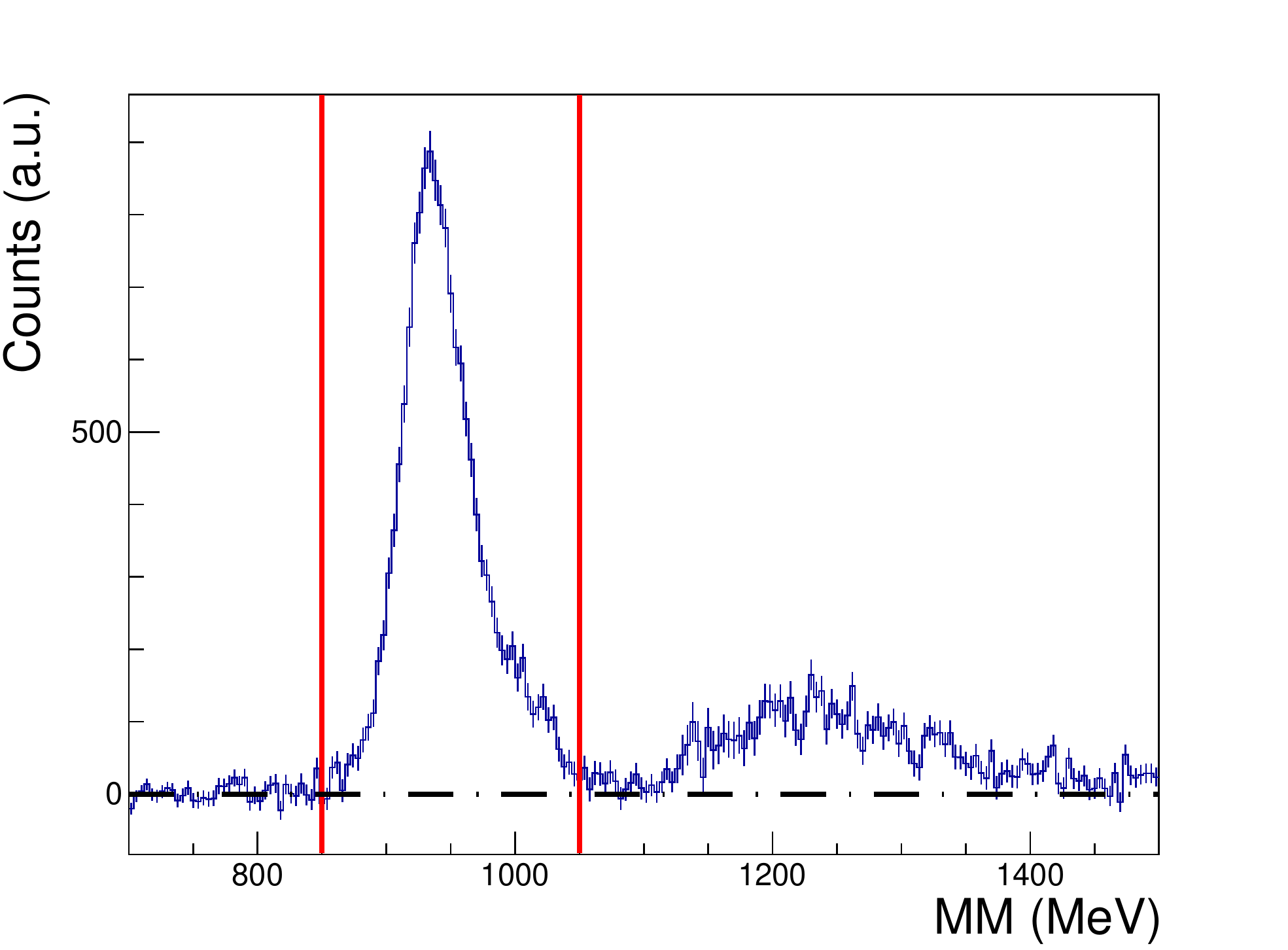}
\caption{Missing mass (MM) distribution for
  the sum (top) and the difference (bottom) of
  events obtained with the parallel and anti-parallel  photon-target
    helicity configurations.
    %
    The acceptance region for the $N\pi^0$ events is inside the red vertical lines at MM~=~850 and 1050~MeV. 
      The  background evident in the right tail
      of the MM distribution  comes from the $\pi^0\pi^{0}$ and
      the $\pi^0\pi^{\pm}$ channels.
}

\label{figure_MM_hel}
\end{figure}


The obtained MM distribution which,  
in comparison to the free-nucleon case
is broadened due to the Fermi motion of the initial-state nucleon, is shown
in the top panel  of Fig.~\ref{figure_MM_hel}. 
   As seen  in  this figure, 
   a consistent background was still present, in particular in the right tail of the distribution. This was mainly due to unpolarised carbon and oxygen nuclei in the butanol molecules. In the bottom part
   of Fig.~\ref{figure_MM_hel}, the difference between the missing mass
   distributions
    of events obtained with the parallel and anti-parallel  photon-target
    helicity configurations
   is shown.

   This 
   allowed a verification of the previous hypothesis since, in this case, the  background from 
   unpolarised nuclei cancels. As expected, the tails become small on both
   sides of the peak and the distribution is centered at the nominal value of the nucleon mass.

Only  events with a MM  between 850 and 1050 MeV (the region between the vertical lines of Fig.~\ref{figure_MM_hel}) were taken into account for the following steps of the analysis.
%
This cut, while accepting most of the $N\pi^0$ events,
eliminates  all the  background showing up on the right tail
of the MM distribution, that  
comes from the $\pi^0\pi^{0}$ and $\pi^0\pi^{\pm}$
processes when the additional photoproduced pion had, at least partially, escaped the detection inside our apparatus. 

\subsection{Proton and neutron identification}

For the evaluation of the $E$ asymmetry for the single $\pi^{0}$ on quasi-free protons and neutrons, 
only  events having one additional 
charged or neutral hit
not used for the $\pi^{0}$ reconstruction were selected from the previously obtained sample.

In the first step of this analysis, the coplanarity distribution between the reconstructed $\pi^{0}$ and the additional track was checked, since, when the Fermi momentum of the target nucleon is neglected, 
the incident photon, the $\pi^{0}$ and the recoil nucleon lie in the same plane, due to momentum conservation.
Simulations showed that the effect due to
    the Fermi motion of the target nucleon does not change the peak position, but only slightly enlarges the width of the distribution.

\begin{figure}%
\centering
\includegraphics[scale=0.38]{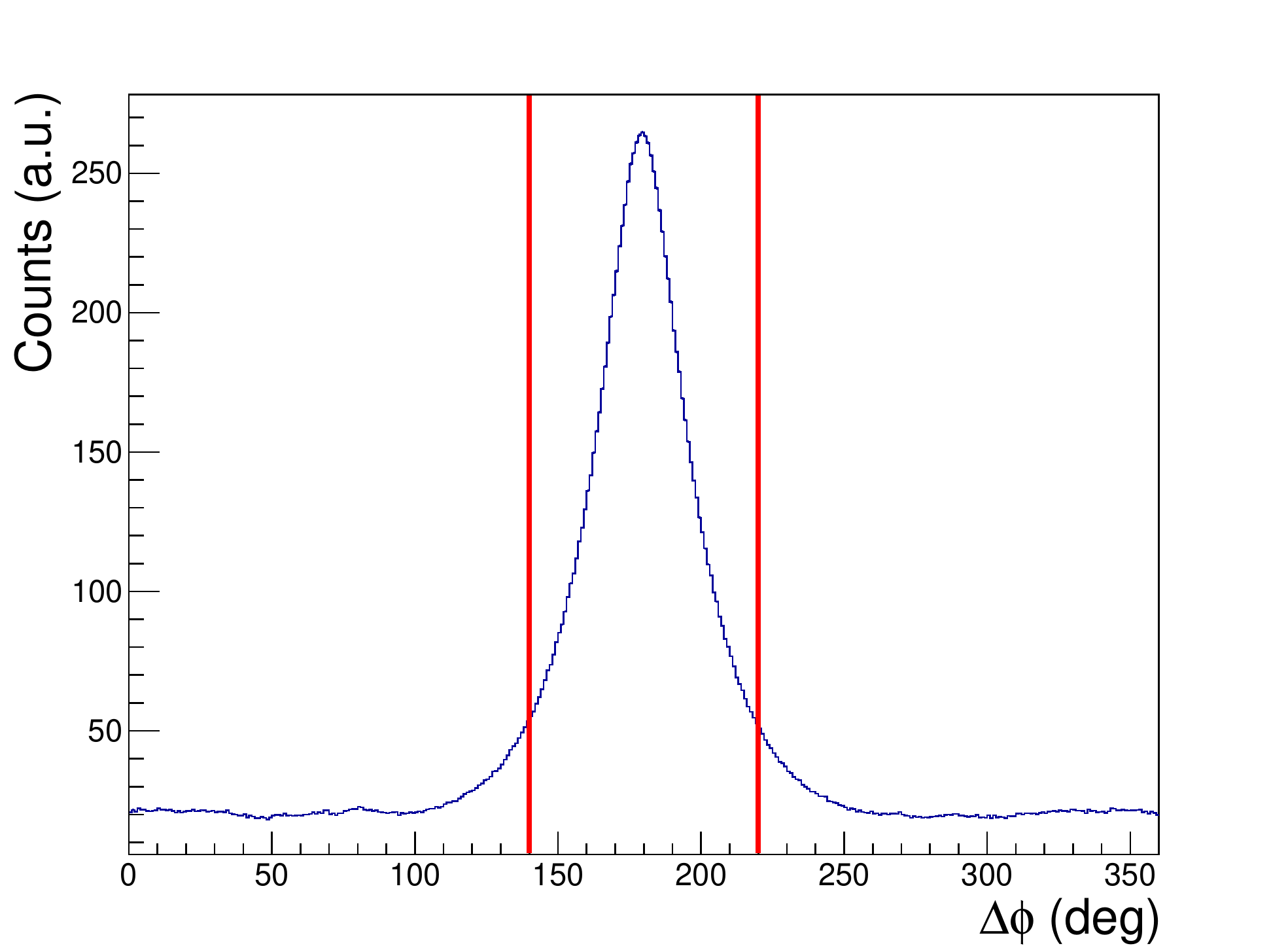}
\caption{Distribution, of the  $\Delta\phi$ azimuthal angle between the reconstructed $\pi^{0}$ and the nucleon track candidate for all events accepted by the MM cut.  The acceptance event region is inside the red vertical lines at $\Delta\phi=140^{\circ}$ and 220$^{\circ}$.}
\label{coplanarity_distribution}
\end{figure}

The mean value of the difference $\Delta\phi$ between the azimuthal angles of the  $\pi^{0}$ and the recoil nucleon must therefore be 
$180^\circ$, as evident from the  $\Delta\phi$  distribution
presented in Fig.~\ref{coplanarity_distribution} for event with a reconstructed $\pi^0$ and
a candidate nucleon track.
%
%
Events having this additional track not co-planar with the identified $\pi^{0}$, i.e. when $\Delta\phi$ was outside  the acceptance region defined in the previous figure, 
were removed from the analysis.
Tracks satisfying both the coplanarity and the MM condition were considered to be proton or neutron candidates depending on whether the track is charged or neutral,
that is with or without a hit either in the PID
or in the VETO detector.

Thereafter additional conditions, discussed below, are applied
to reject charged (neutral) tracks that could be misidentified as protons (neutrons).

In the forward region covered by TAPS, it was possible to perform a Pulse Shape Analysis (PSA)~\cite{diet1}, thanks to the two ("fast" and "slow") components of the signals from the BaF$_2$ crystals. These components were integrated over two different ranges
(short gate: 40~ns; long gate: 2$~\mu$s) to obtain the $E_s$ and $E_l$ energy components, respectively. 
For photons, the two components are quite similar, while, for massive particles $E_s$ is smaller than $E_l$. To better highlight this difference, it is convenient to use the transformation to the PSA radius $r_{PSA}$ and angle $\phi_{PSA}$, which are defined as:
\begin{equation}\label{eqpsa}
 r_{PSA} = \sqrt{E_s^2 + E_l^2} \quad ; \quad \phi_{PSA}=\arctan{E_s/E_l} \ .
\end{equation}
    Since, for photons $E_s \approxeq E_l$,
    while, for massive particles  $E_s < E_l$, photons are evident
    at  $\phi_{PSA} \approxeq 45^\circ$, independently of $r_{PSA}$, 
    while neutrons are located at smaller angles.

\begin{figure}%
\centering
\includegraphics[scale=0.35]{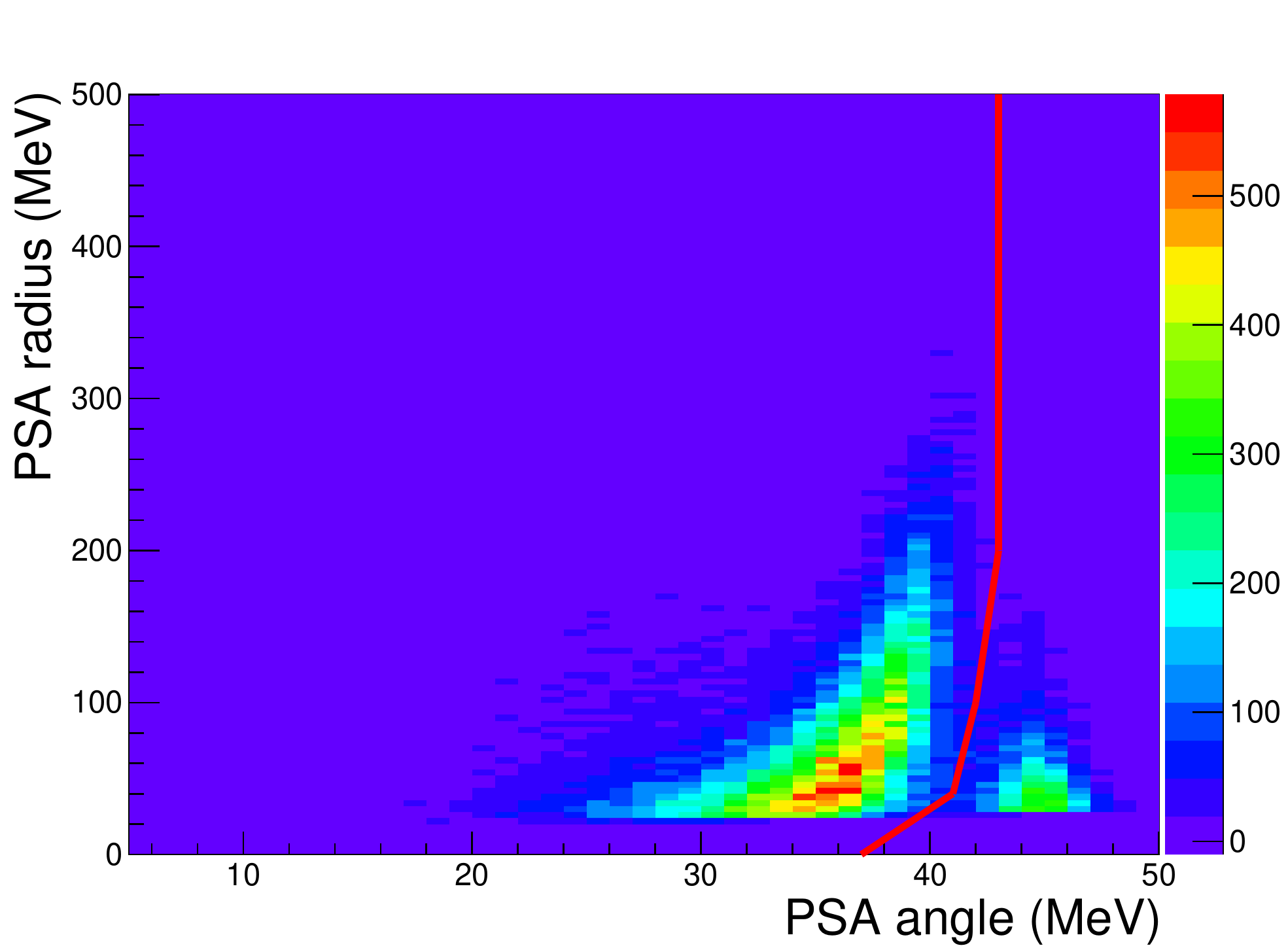}
\includegraphics[scale=0.35]{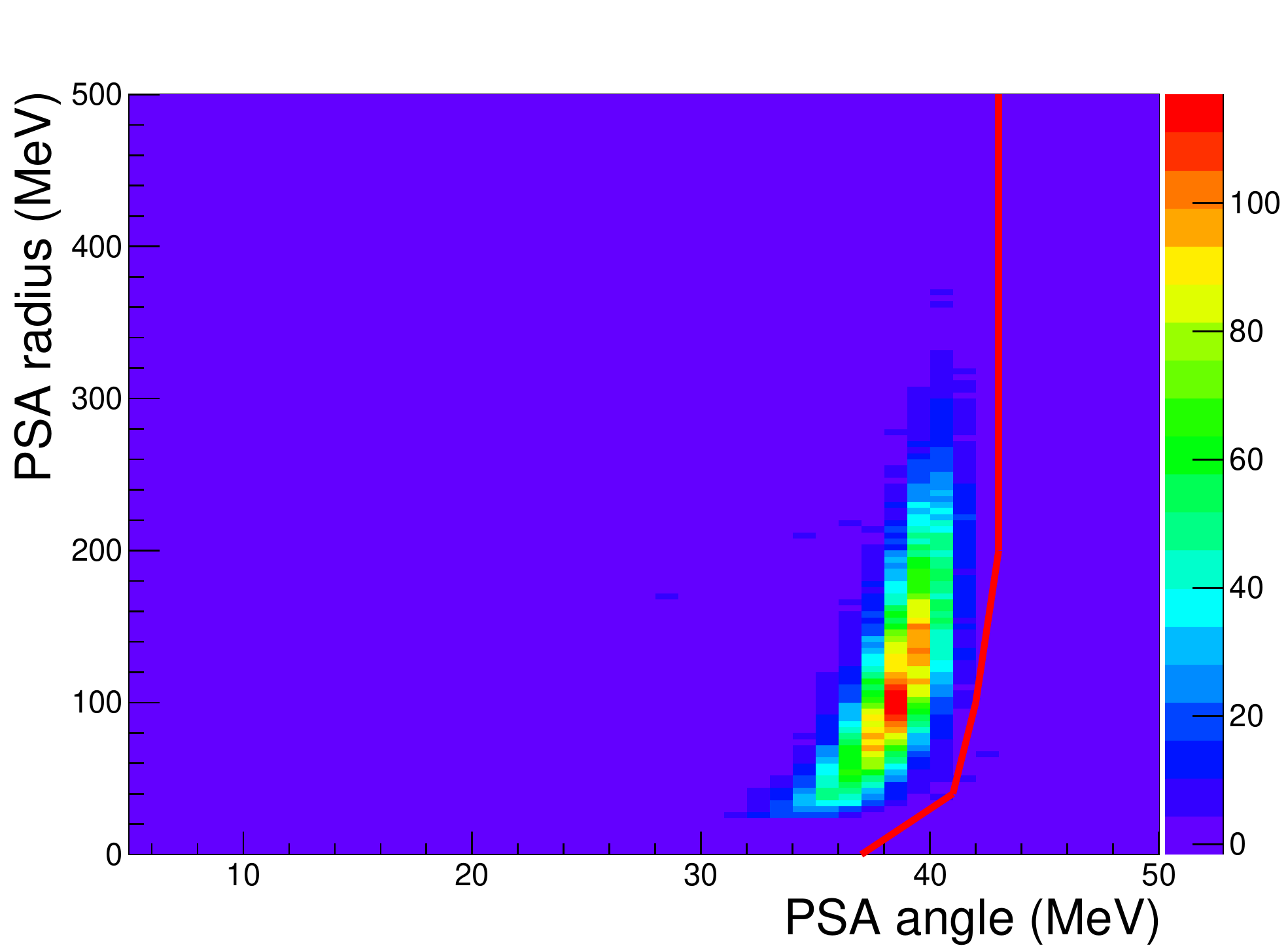}
\caption{PSA plots for neutral (top) and charged (bottom) tracks after $\pi^{0}$ reconstruction, MM and coplanarity cuts.
  The neutron and proton candidates lying on the right of the red lines have been rejected.
}
\label{PSA_charged}
\end{figure}

In Fig.~\ref{PSA_charged}, the obtained PSA spectra for the proton and neutron candidates are given. Events with  particle candidates on the right of the red curve were rejected. 
In the charged track case, no relevant background was present even before this cut was applied.

    Due to the good time resolution of the TAPS detector and the relatively long distance between the target and the detector (about 1.5~m), a time-of-flight (ToF) analysis was also performed,  to refine both the neutron
    and the proton selection.
    
    In this case,  photon candidates formed a band at a constant ToF
    corresponding to the target - detector, distance
    while non-relativistic protons and neutrons were located in a band at
    higher ToF values.
    
\begin{figure}[ht]
\centering
\includegraphics[scale=0.35]{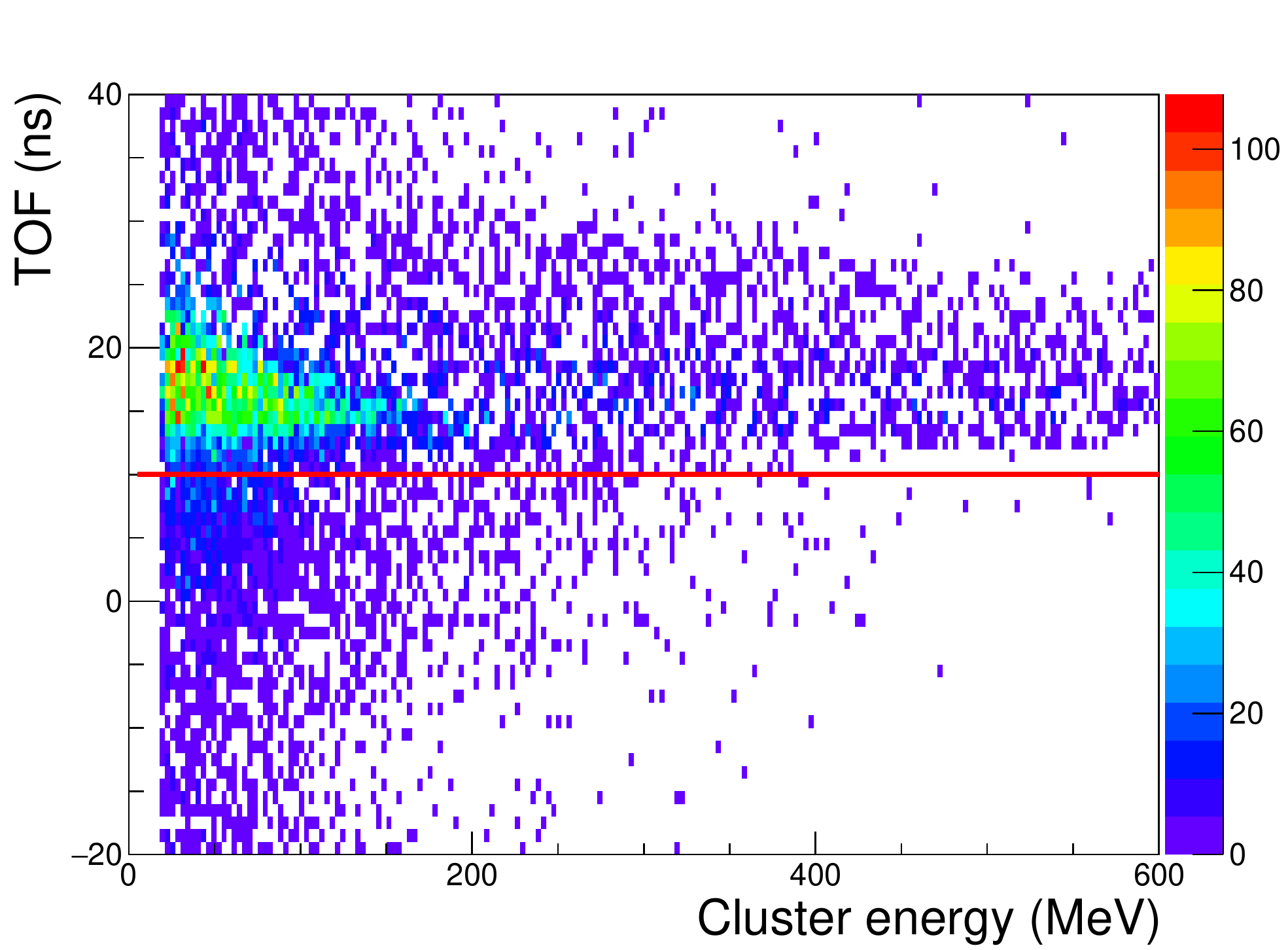}
\includegraphics[scale=0.35]{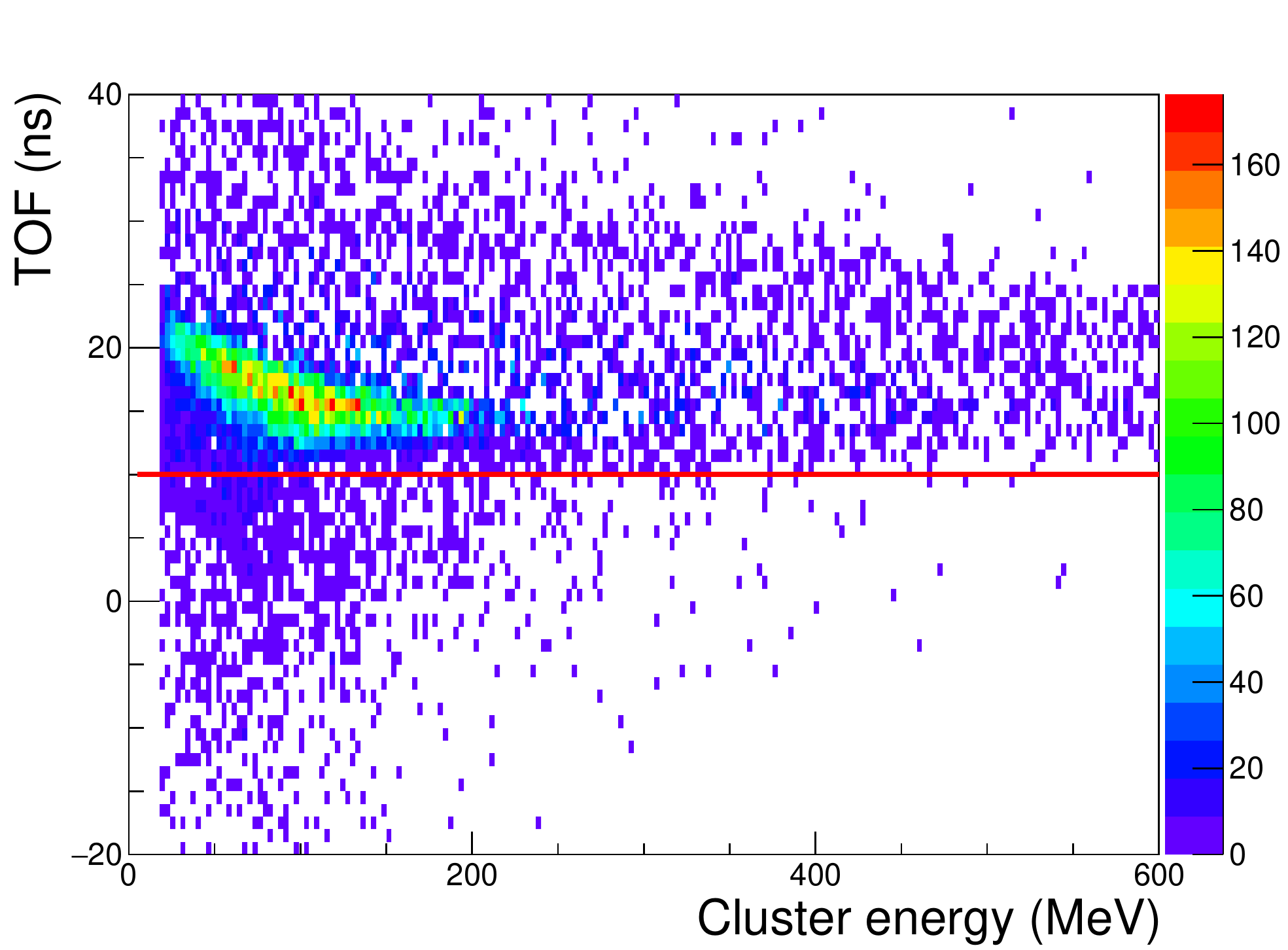}
\caption{ToF analysis for neutral (top) and charged (bottom)
  tracks after the PSA cut.
  The neutron and proton candidates lying below the red lines have been rejected.}
\label{TOF_neutral_cut}
\end{figure}

    The results of this analysis are shown in Fig.~\ref{TOF_neutral_cut}, where the ToF (expressed as difference with the event trigger time) of
    both for the accepted neutron (top plot) and proton (bottom plot) candidates after the PSA cut
    is compared to the total deposited particle energy.

    Guided by the simulation, a residual background was rejected by the
    horizontal red lines shown in  Fig.~\ref{TOF_neutral_cut}.
    
For neutral particles detected in the CB, a cluster size analysis was used to
cross check  the  neutron selection performed with
the $\chi^2_{\pi^0}$ selection method previously described (see Eq. \ref{eqchi2}).
%

As shown by a simulation of the quasi-free $n\pi^0$ process,  
neutron clusters consisted of very few detector elements
(just one in many cases),
while high energy photons coming from the $\pi^0$ decay produced, on average, larger clusters due to the much
 larger amount of deposited energy. 
 
\begin{figure}[ht]
\centering
\includegraphics[scale=0.35]{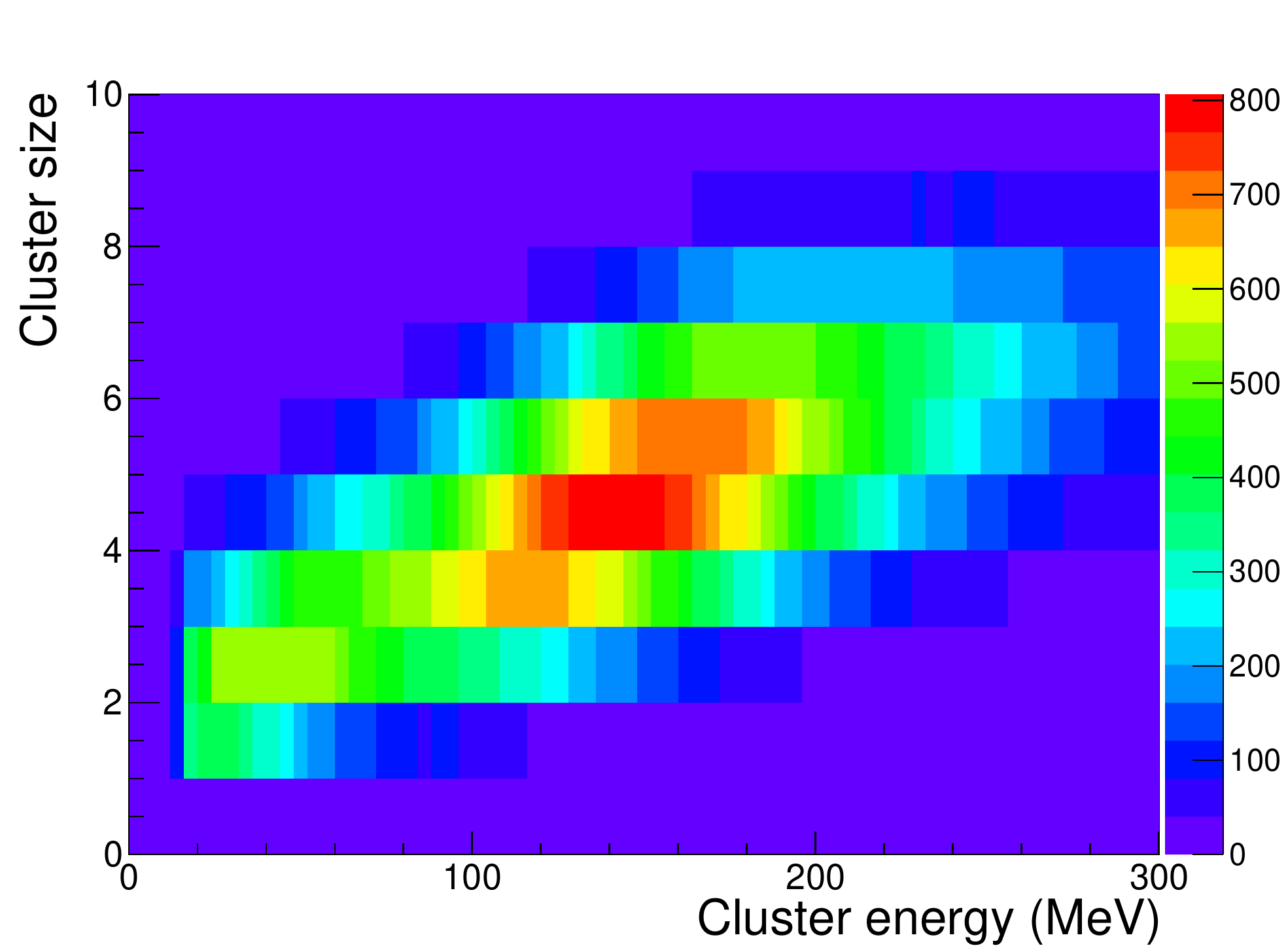}
\includegraphics[scale=0.35]{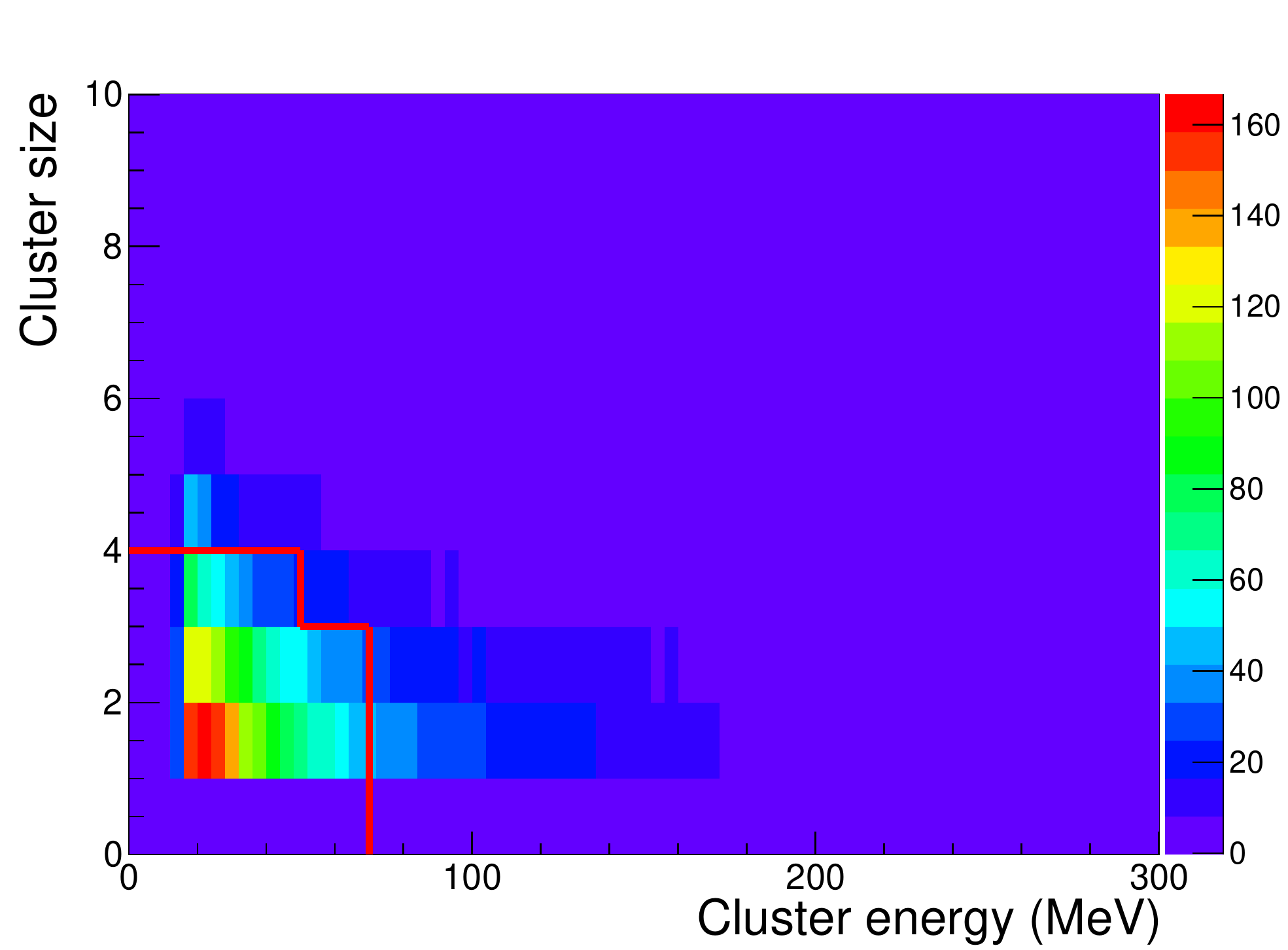}
\caption{
  Cluster size distribution versus the energy released in CB
  for all tracks of events with 2 neutral clusters and a reconstructed $\pi^0$ (top plot) and for the unapaired clusters in events with 3 neutral clusters and a reconstructed $\pi^0$ (bottom plot).
  The tracks lying below the red line were considered as neutrons.
    }
\label{cbcluster}
\end{figure}

 In Fig.~\ref{cbcluster}, the experimental cluster size distribution
 is compared to the deposited energy in CB
for photons coming from the  $\pi^0$ decay,  
selected from events with two neutral clusters and a reconstructed $\pi^0$ (top plot),
and for the third cluster, not selected as part of a $\pi^0$ decay,
in events with 3 neutral clusters (bottom plot).
%
Neutrons from the $n\pi^0$ channel
congregate in the bottom-left part of the plot, while photons mainly
populate the mid and top-left parts.
Guided by the quasi-free $n\pi^0$ simulation, a final selection cut,
shown by the red line in the
bottom plot of Fig.~\ref{cbcluster},
was applied so that  
no significant background  is left after the end of the neutron
selection procedure.


\begin{figure}[ht]
\centering
\includegraphics[scale=0.36]{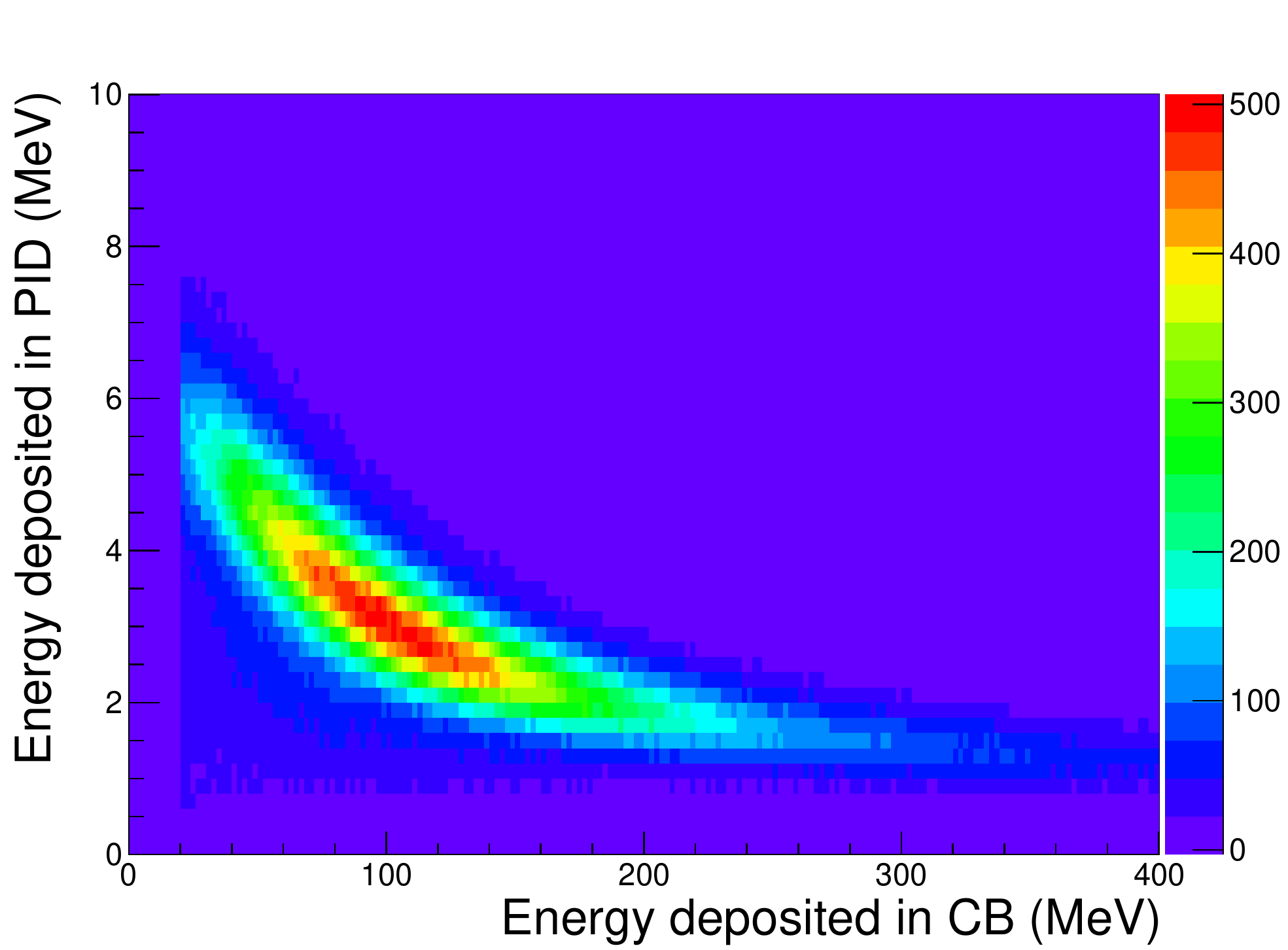}
\includegraphics[scale=0.36]{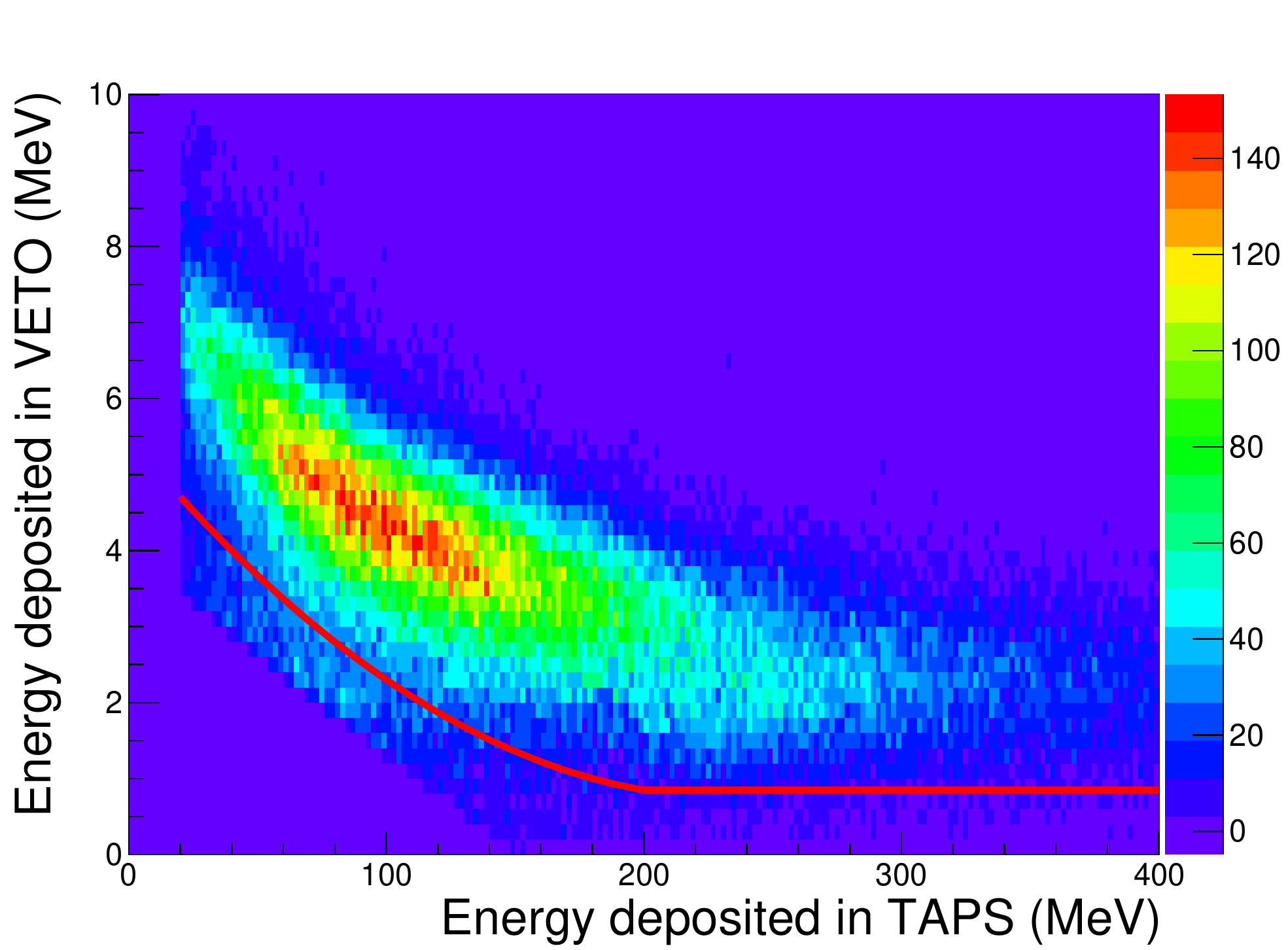}
\caption {$\Delta E-E$ plot with the energy information from PID and CB
  (top plot) and VETO and TAPS (bottom plot)  after coplanarity and MM cuts.
 In the bottom plot, tracks lying above the red line were considered as protons.
}
\label{dE_CB_PID}
\end{figure}

As a final cross check of  the proton selection analysis, the 
$\Delta E-E$ plots both for tracks detected in CB (using PID and CB energy information) and TAPS (using VETO and TAPS energy information) were constructed. 
    Figure~\ref{dE_CB_PID} shows the $\Delta E-E$ plot obtained with the PID-CB
    and the VETO-TAPS detectors, respectively.
    In both cases, the
    proton band is very clean, which proves the validity of the selection procedure.
    For the VETO-TAPS combination, a final selection cut, shown by the red line  in the bottom plot of  Fig.~\ref{dE_CB_PID}, was applied to suppress a small residual background.

\subsection{Unpolarised background subtraction}\label{unpol}
In the extraction of the $E$ observable, the evaluation of the background coming from unpolarised C and O target  nuclei was crucial for the correct evaluation of the denominator of Eq.~\ref{E_obs2}.
As previously mentioned, dedicated data  were taken with a carbon foam target
to separately measure this background contribution, 
under the assumption that the nucleons bound in C and O
nuclei give the same response to the incoming photons.
 
Due to this effect, Eq.~\ref{E_obs2} has to be modified as:

\begin{eqnarray}\label{eq-E-obs}
&&E_{p}(W,\theta)=\frac{1}{P_{z}^{T}}\cdot\frac{1}{P_{\odot}^{\gamma}}\\
&&\phantom{xxx}\times\frac{N_{\mathrm{BUT}}^{\uparrow\downarrow}
(W,\theta)-N_{\mathrm{BUT}}^{\uparrow\uparrow}(W,\theta)}
{(N_{\mathrm{BUT}}^{\uparrow\downarrow}(W,\theta)+N_{\mathrm{BUT}}^{\uparrow\uparrow}(W,\theta))-s\cdot N_{C}(W,\theta)},\nonumber
\end{eqnarray}
where $W$ in the total center-of-mass energy and 
the subscripts "BUT" and "C" indicate the data from butanol and carbon targets, respectively, and $s$ is the scaling factor needed to normalize the different data sets.

 The scaling factor $s$ was determined using different methods: a)
 absolute normalisation by photon flux, target density and detection efficiency; b) using MM or coplanarity spectra in a region where the quasi-free nucleons  do not contribute (MM~$\sim 1050$~MeV). 

\begin{figure}%
\centering
\includegraphics[scale=0.45]{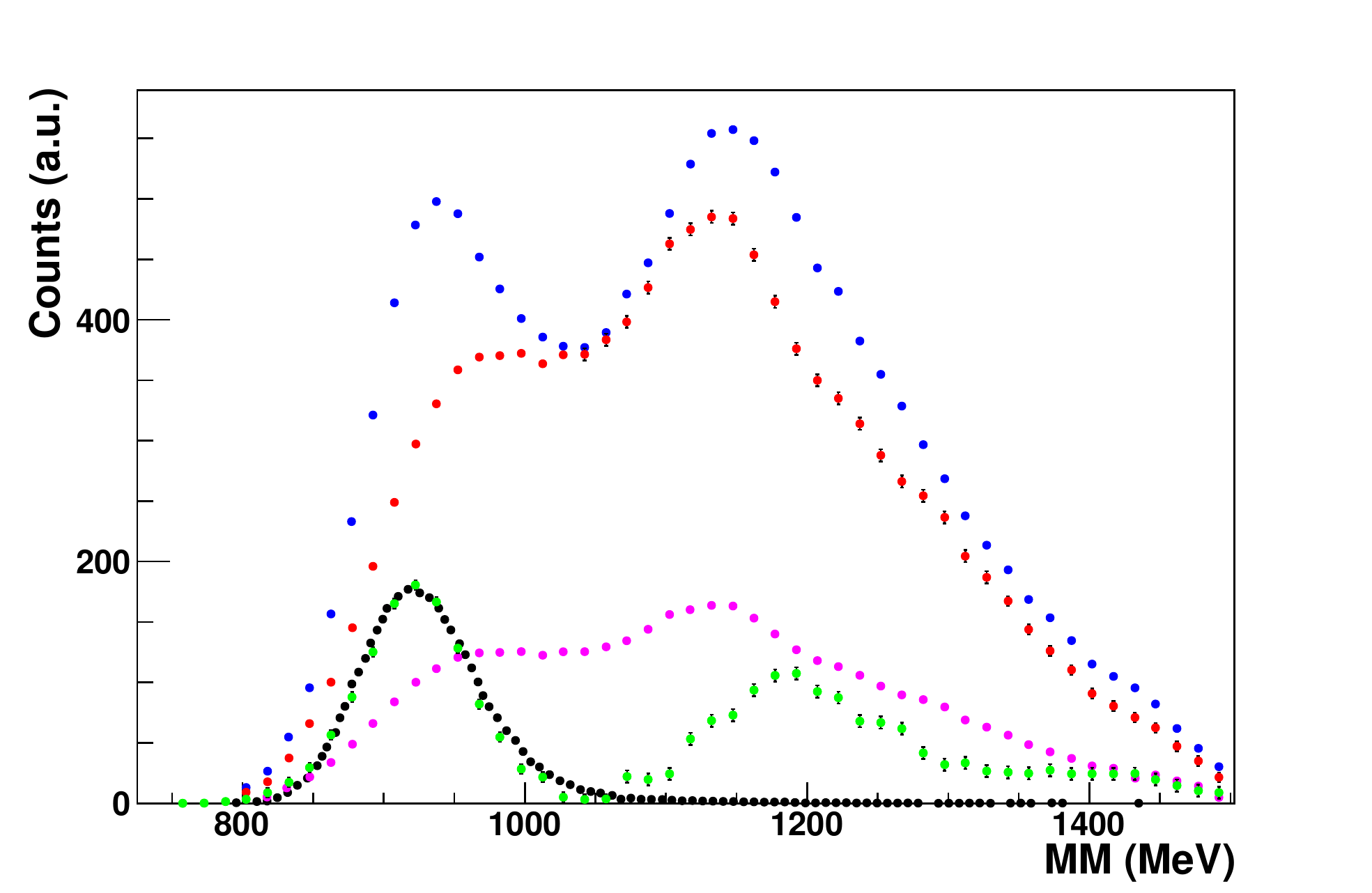}
\caption {Subtraction of the carbon background for proton events using the missing mass plots obtained from D-butanol and carbon data for $W$ (total center-of-mass energy)
  between 1450 and 1480~MeV. The different points represent the missing mass distribution for: the D-butanol target (blue dots), the carbon target (magenta dots), the scaled carbon (red dots) and the proton events (green dots) obtained by subtracting the scaled carbon events from the D-butanol events. The black dots show the simulated distribution of quasi-free proton events from a deuterium target.}
\label{scaling_sketch}
\end{figure}

     A typical example of the obtained MM spectra is shown in Fig.~\ref{scaling_sketch} for events with quasi-free protons.
    The $s$ factor was used to scale the original carbon distribution (magenta dots  in Fig.~\ref{scaling_sketch}) and the MM distribution from quasi-free protons bound inside the deuteron (green dots) was evaluated
    by subtracting the scaled carbon distribution (red dots) from the one from deuterated butanol (blue dots).
The subtracted distribution
is in very good agreement with  the simulated quasi-free proton distribution
(black dots) for MM values below 1050 MeV.  
As in Fig.~\ref{figure_MM_hel}, it also shows the good rejection of events
from double pion reactions achieved with the MM cut.
     In general, the more pronounced unpolarised contributions
    were found at the highest photon energy  
values and in the most extreme angular regions.
 %
%

\begin{figure}[ht]
\centering
\includegraphics[scale=0.35]{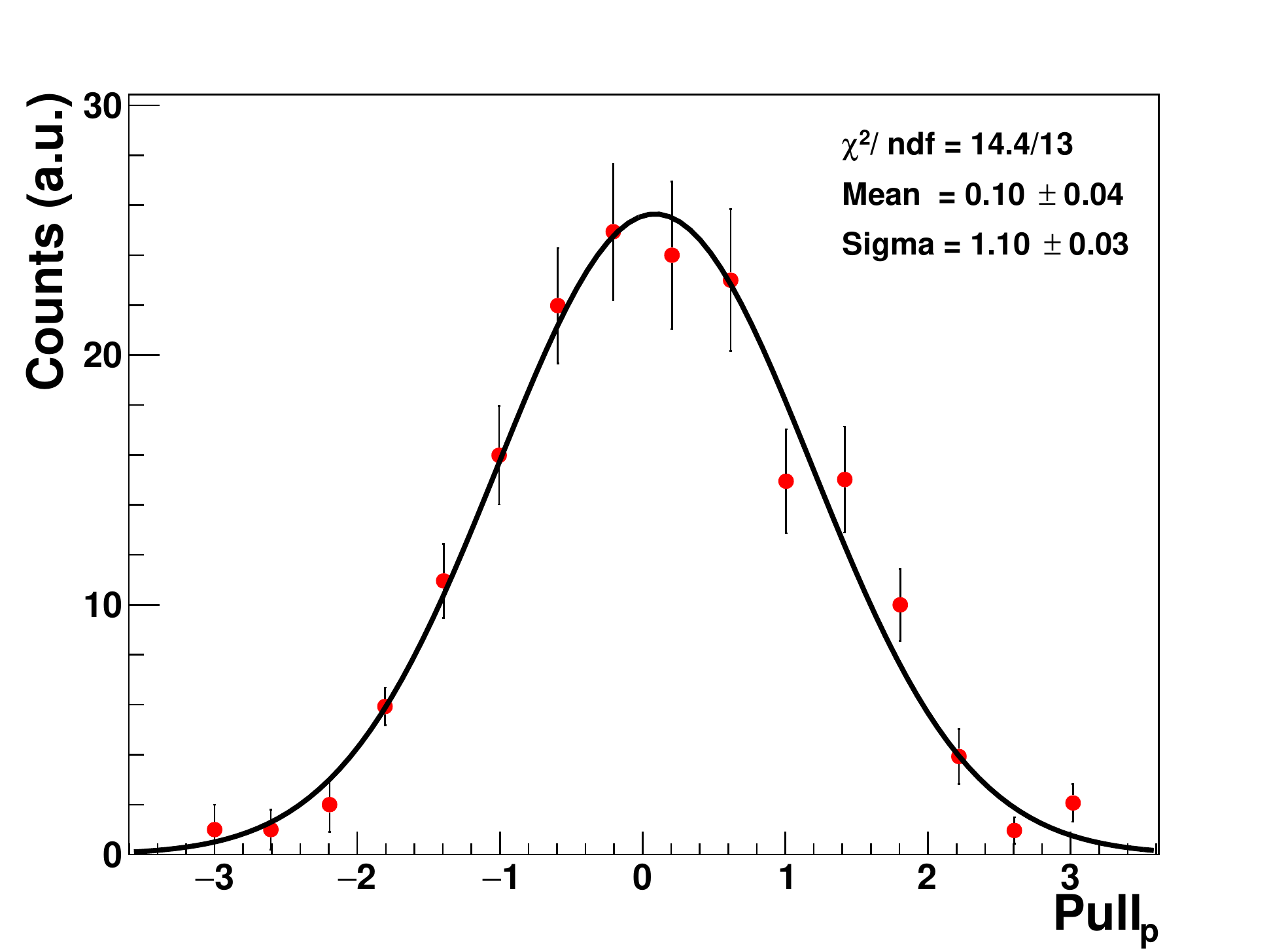}
\includegraphics[scale=0.35]{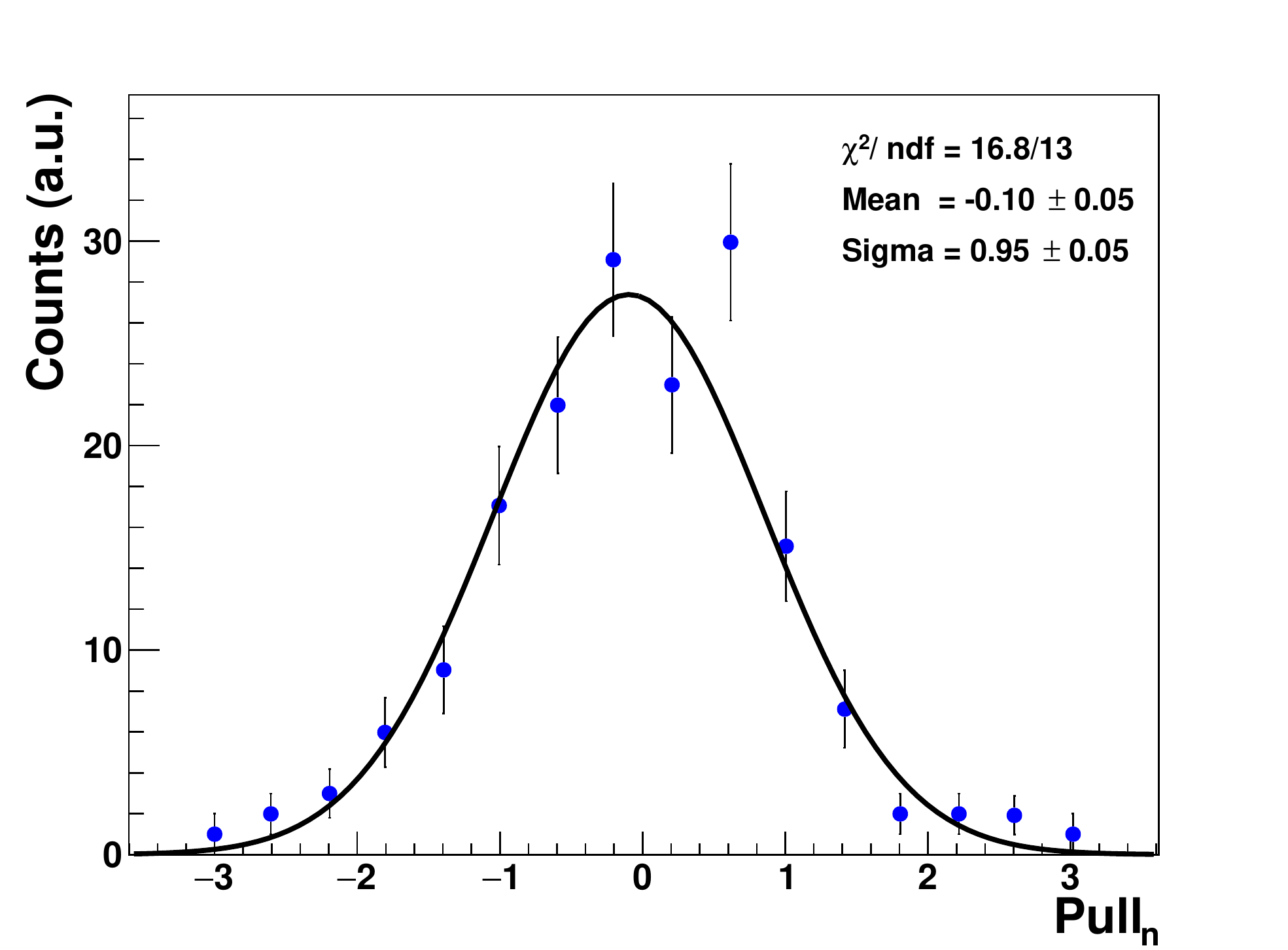}
\caption{Distribution of the Pull variable (Eq.~(\ref{eqpull}))
  for protons (top) and neutrons (bottom).}
\label{figpull}
\end{figure}

    The two methods a) and b) described above gave quite similar and
    statistically equivalent results. As an example,
    in Fig.~\ref{figpull} the distribution of the Pull variable:
    \begin{equation} \label{eqpull}      
            Pull_{p(n)}= \frac{(E_1p(n) -E_2p(n))}
      {\sqrt{\sigma_{1p(n)}^2+\sigma_{2p(n)}^2}}
      \end{equation}
   is shown,  where $E_1p(n)$ and $E_2p(n)$ are the asymmetries evaluated at each $\theta$ and $W$   value using these two methods 
on different data subsets for the proton (neutron) case.
The solid lines represent the best-fit gaussians obtained from the data,
whose parameters are given in the legends.
According to expectations, both the mean and the variance resulting from
the fit are compatible with the standard gaussian parameters. 
The final $E$ central values are taken as the weighted average between the different procedures~\cite{fede,manu}.

The differences in the $E$ values obtained with these methods were used to evaluate the systematic uncertainties associated with this procedure. In most of the measured energy and angular bins, the relative values of these uncertainties were estimated to be in the range  $2\%$ to $10\%$, with the exception of the most forward angular bin, where the limited statistics due to the low detector efficiency often causes larger uncertainties.

This procedure was  performed independently for proton and neutron events, with different scaling factors obtained for each $W$ bin, when method b) was applied.
The angular dependence of the scaling factors was also checked, but found to be negligible.

\subsection{Inclusive single $\pi^{0}$ photoproduction on the deuteron: $\gamma d \to \pi^0 B \ (B=pn\, \hbox{or}\, d)$}
%
The helicity-dependent cross section difference  
$({{d}\Delta\sigma}/{{d}\Omega})$
for single $\pi^{0}$
on the deuteron can be expressed as follows (see also Eq.~\ref{E_obs2})  :
\begin{eqnarray}\label{eq:beam-recoil}
&&\frac{\mathrm{d}\Delta\sigma}{\mathrm{\mathrm{d}\Omega}}(E_{\gamma})
=\frac{\mathrm{d}\sigma^{\uparrow\uparrow}}{\mathrm{\mathrm{d}\Omega}}(E_{\gamma})
-\frac{\mathrm{d}\sigma^{\uparrow\downarrow}}{\mathrm{\mathrm{d}\Omega}}(E_{\gamma}) =\\
&&\phantom{xx} =2\cdot\frac{N^{\uparrow\uparrow}(E_{\gamma},\theta)-N^{\uparrow\downarrow}(E_{\gamma},\theta)}
{I_{\gamma}(E_{\gamma})\cdot\epsilon_{DET}(E_{\gamma},\theta)\cdot\Delta\Omega\cdot n}\cdot\frac{1}{P_{\odot}^{\gamma}}\cdot\frac{1}{P_{z}^{T}},\nonumber
\end{eqnarray}
    where $I_\gamma$ is the total photon flux, with
    $I_{\gamma} = 2 I^{\uparrow\uparrow}_\gamma = 2 I^{\uparrow\downarrow}_\gamma $
       due to our experimental conditions (see Sect.~\ref{expapp}),
%
       $\epsilon_{DET}$ is the 
       $\pi^0$ reconstruction efficiency, as determined by the GEANT4 simulation,
$\Delta\Omega$ is the solid angle factor
       and $n$ is the surface density of polarised de\-u\-te\-rons.
For this observable, it was necessary to select all events with a $\pi^{0}$ reconstructed in the CB-TAPS setup, without additional requirements.
%
The relative systematic uncertainty of $\epsilon_{DET}$, estimated to be $4\%$, was evaluated
by examining the cross section variations due to the  
different cuts and selection conditions applied both to the experimental and the simulated data. 
    The values of the helicity-dependent total cross section difference
    $\Delta\sigma$ 
    were obtained by
integrating Eq.~(\ref{eq:beam-recoil}) over the full solid angle. 
 In this case, no unpolarised contribution needed to be evaluated since the effect of the unpolarised C and O spinless nuclei in the target vanish in the difference. 

\subsection{$E$ asymmetry for single $\pi^{0}$ on quasi-free protons and neutrons}

In addition to the detection of one $\pi^{0}$, the events selected during the analysis were required to also have a proton or neutron identified. A good quality of the nucleon selection from any polarised and unpolarised background is  crucial for a highly precise calculation of the $E$ observable. This goal has been achieved by 
the selection previously described.

\subsection{Systematic uncertainties} \label{par:syserr}

The different sources of systematic uncertainties previously discussed are summarized in Tab.~\ref{tab:syserr}.

Sources of common global
 systematic uncertainties  come from the
 absolute photon flux normalization, particle reconstruction efficiency (these contributions are only relevant for the cross section evaluation), from  the beam and target polarisation values and from  the target surface density.

 The point-to-point systematic uncertainty contribution  from the
 unpolarised background subtraction is only relevant for the $E$ observable and it is dependent on the analysed $W$ and $\rm{cos}(\theta^{CM}_{\pi^0})$ bins, as described in Section~\ref{unpol}.

\begin{table}[ht]
\centering{}
\captionsetup{width=0.95\linewidth}
\caption{Systematic uncertainties of the present data analysis. \label{tab:syserr}}
\begin{tabular}{|l|c|}
  \hline
  Target polarisation & $\pm 10\%$ \\
  Unpolarised background subtraction& $\pm 2-10\%$ \\
  Tagging efficiency & $\pm 4\%$ \\
  Detector efficiency & $\pm 4\%$ \\
  Beam polarisation & $\pm 3\%$ \\
  Target filling factor & $\pm 2\%$ \\
   \hline
  \end{tabular} 
\end{table}

\section{Results}

\subsection{Inclusive single $\pi^{0}$ photoproduction cross section on the deuteron}

\begin{figure*} %
  \centering
  \includegraphics[scale=0.75]{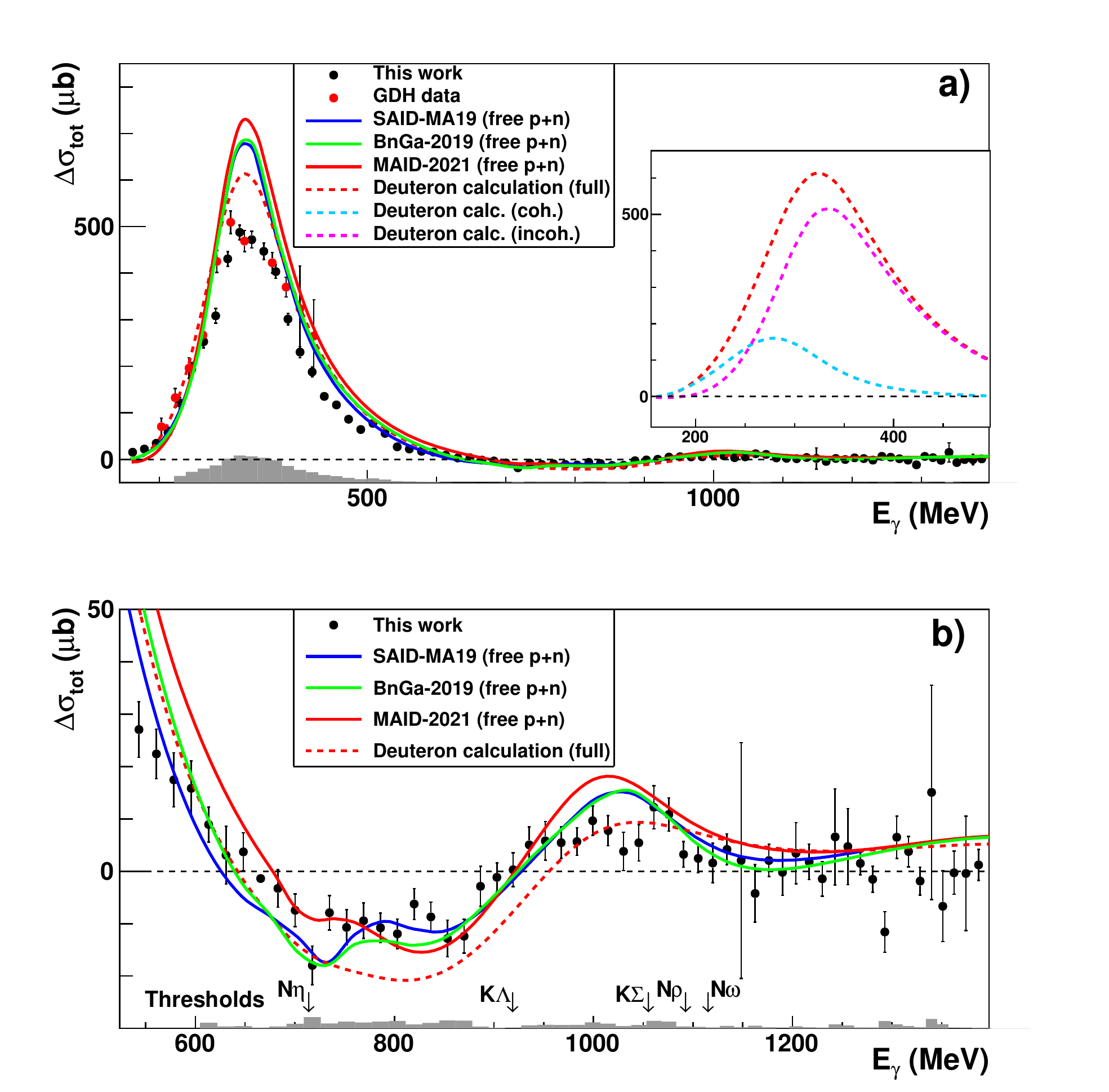}
%
%
\caption{Inclusive polarised $\pi^{0}$ photoproduction cross section on the deuteron
  ($\gamma d \rightarrow \pi^{0}B$).
  The new results (black points) are compared with the results from the GDH Collaboration (red points)\cite{ahr09}.
  The different solid lines  show the predictions for the free
(proton+neutron) sum of different analyses. Blue line: 
SAID-MA19~\cite{SMA19} ; green line:
BnGa-2019~\cite{boga};  red line:
MAID-2021~\cite{victor}. 
The dashed line shows the predictions for $\gamma d \rightarrow \pi^{0}B$  ($B=pn\, \hbox{or}\, d$)  
obtained using the model of Refs.~\cite{fixar1,fixar2} with the new MAID-2021 amplitudes for $\gamma N\to\pi N$.
The insert shows separate contributions from the incoherent $\gamma d \rightarrow \pi^{0} pn$
and the coherent  $\gamma d \rightarrow \pi^{0}d$ channels. In (b)
the markers at the energies corresponding to opening of other channels are also indicated.
The contributions of all
the systematic uncertainties (see Sect.~\ref{par:syserr})  are depicted as grey bars.
 }
\label{total_pol_cs}
\end{figure*}


The total  helicity-dependent cross section difference $\Delta\sigma$ 
for the 
$\gamma d \to \pi^{0} B\, (B=pn\,\hbox{or}\, d)$ reaction,
is shown in Fig.~\ref{total_pol_cs}a)
(black points) in the region from E$_\gamma$=160~MeV up to 1390~MeV.
It is compared to the data (red points) previously published by the GDH collaboration~\cite{ahr09}.
With respect to the previous results, this
work provides new data covering a wider energy range with better statistics.

In Fig.~\ref{total_pol_cs}b),
only the results for E$_{\gamma}$ > 550 MeV are plotted to better
highlight the high-energy behaviour.
The different solid lines show the predictions for the elementary
(proton+neutron) cross sections given by different multipole analyses:
SAID-MA19~\cite{SMA19} (blue line);
BnGa-2019~\cite{boga} (green line);
MAID-2021~\cite{victor} (red line).

%
    All these analyses use  coupled-channel
    approaches to derive the different multipoles from the available experimental
    database, but differ in the parameterisation of the resonant and background parts of the
    photoproductions amplitudes and in the treatment of the constraints
    (like unitarity, analyticity, gauge invariance, crossing and chiral simmetries) imposed
    by general theoretical considerations (for more details, see Ref.~\cite{allPWA}).

The discrepancy between the experimental results and the free (proton+neutron)
calculation is primarily due to  nuclear effects, in particular FSI,
which are especially important in the $\Delta(1232)$ resonance region
(150~MeV~$\lesssim E_{\gamma} \lesssim$~500~MeV).
The dashed red line shows the theoretical results given by
    the calculation performed by A.Fix,
    which takes into account nuclear effects in the deuteron.
    It is based on
    the model of Refs.~\cite{fixar1,fixar2}, in which the elementary
    $N\pi$ amplitudes are embedded in the deuteron wave function and
    FSI are incorporated in a perturbative manner.
    For this prediction,
    amplitudes for the $\gamma N\to\pi^0 N$  channel from 
        MAID-2021 were used
        instead of the MAID-2003 version used in~\cite{fixar2}.

In the insert  of Fig.~\ref{total_pol_cs}a)
the predictions of the coherent ($\gamma d\to\pi^0 d$) and incoherent ($\gamma d\to\pi^0 pn$) cross sections, calculated  using the model of A. Fix,
are also presented separately. It  can be seen that the coherent process gives a sizeable contribution to the $\pi^0$ production process only at photon energies below 400~MeV.
As seen in Fig.\,\ref{total_pol_cs}, after the nuclear effects are included,
the calculated cross section difference $\Delta\sigma$ visibly decreases.
The major source of this reduction, as discussed in~Ref.\cite{fixar2},
is the interaction between the final nucleons in the $^3S_1$ state.
In contrast to the charged channels, $\gamma d\to\pi^+ nn$ and $\gamma d\to\pi^-pp$, the plane wave cross section for $\gamma d\to\pi^0np$ effectively contains a spurious contribution from the coherent channel $\gamma d\to \pi^0 d$. The latter is due to the trivial fact that the final $NN$ plane wave is not orthogonal to the deuteron ground state. After this spurious contribution is projected out, the resulting interaction effect turns out to be of the same order as for charged pion production, about 2\,$\%$. Inclusion of $\pi N$ rescattering leads to a further reduction of the cross section.

\begin{figure*} %
\centering
  \hspace{-1.5cm}
  \includegraphics[scale=0.85]{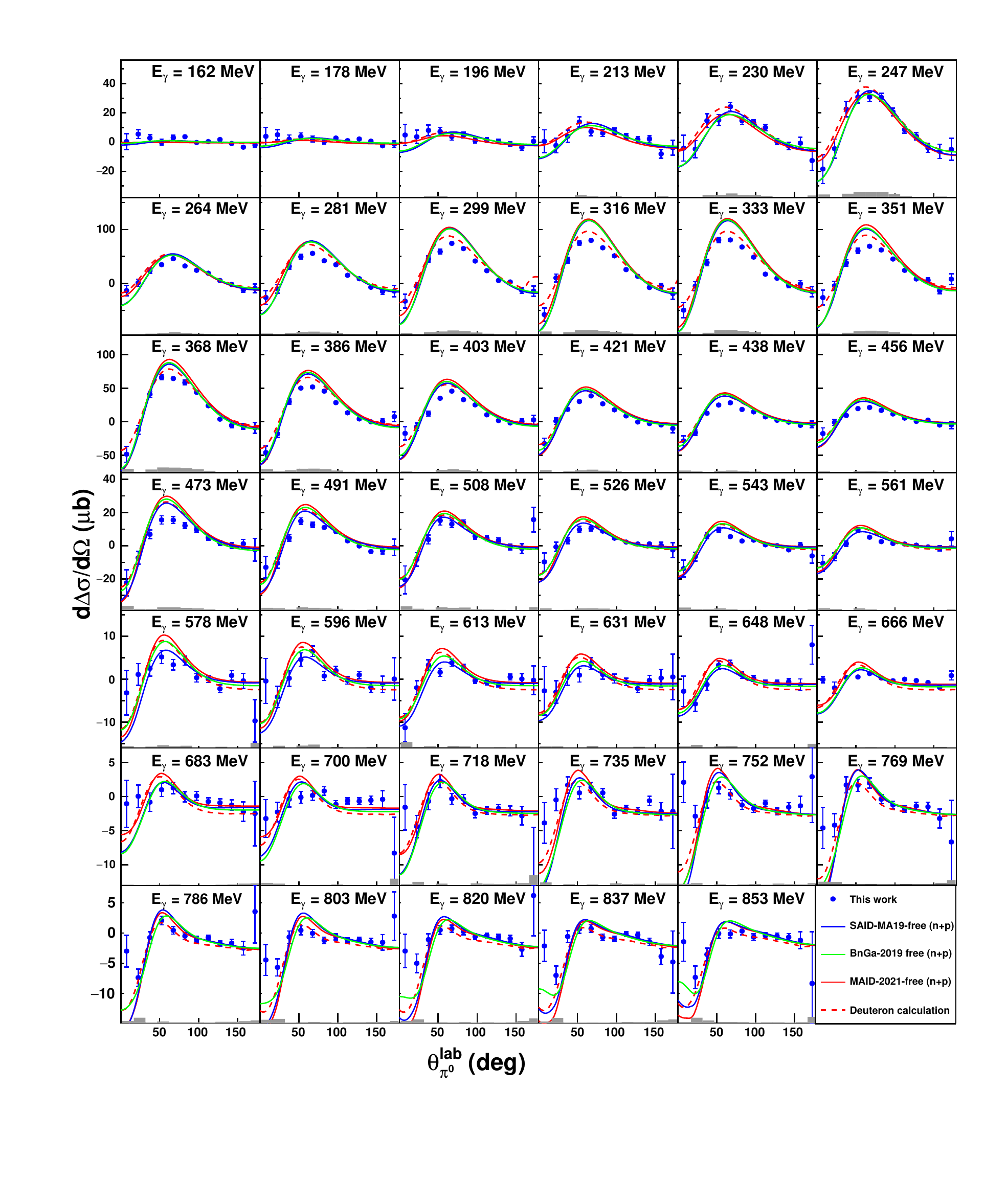}
  \vskip -1.75cm
\caption{Helicity-dependent differential cross section for single $\pi^{0}$
photoproduction on the deuteron as function of $\theta_{\pi^{0}}^{\mathrm{LAB}}$ for $E_{\gamma} <860$~MeV.
The color code for the theory curves is as in Fig.\,\ref{total_pol_cs}.
The contributions of all
the systematic uncertainties (see Sect.~\ref{par:syserr})  are depicted as grey bars.
}
\label{Diff_cross_sec_low_en}
\end{figure*}

\begin{figure*} %
\centering
    \hspace{-1.5cm}
  \includegraphics[scale=0.85]{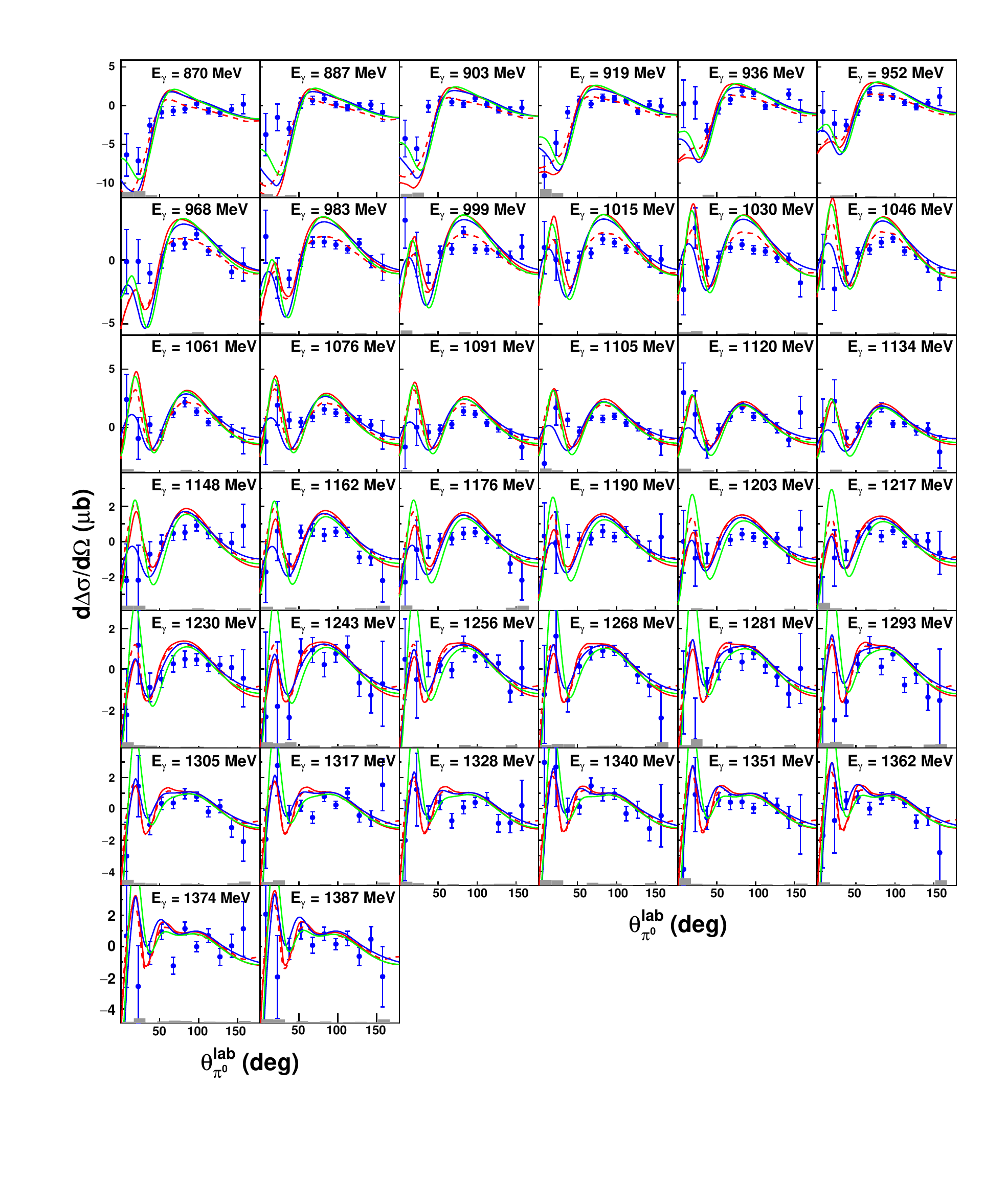}
  \vskip -1.75cm
\caption{Same as in Fig.\,\ref{Diff_cross_sec_low_en} 
for $E_{\gamma}>860$ MeV.
}
\label{Diff_cross_sec_high_en}
\end{figure*}


Thus, the total FSI effect in the $\Delta$ region is a decrease of the total cross section
difference $\Delta\sigma$ by about 20\,$\%$ in both helicity states. At the same time, as evident from Fig.~\ref{total_pol_cs}, this reduction is not sufficiently strong to reproduce the experimental data. The source of the remaining deviation is still unclear. In particular, as shown
in Ref.~\cite{fixar2}, the multiple scattering corrections in the $\pi NN$ system are insignificant, and their inclusion cannot explain the discrepancy.

In  Fig.~\ref{total_pol_cs}b),
as in the free-nucleon case,
a dip in the experimental $\Delta\sigma_{\pi^0}$ values
can  be observed 
near the $\eta$ production threshold
    due to the intereference between $N\pi^0$ and
    $N\eta$ channels,
    while, for E$_\gamma$ values just above $\approx~1$~GeV, the effects due to the excitation of the $F_{15}(1680)$ resonance,
which has both a large value of the helicity amplitude $A_{3/2}$
   and a large $N\pi$ decay branching ratio (see Ref.~\cite{ref:PDG}), 
    are reduced with respect to the free-nucleon case.



The differential cross section difference $d\Delta\sigma / d\Omega$ results
for individual photon energy bins from 162 to 1387~MeV
are shown in Figs.~\ref{Diff_cross_sec_low_en} and~\ref{Diff_cross_sec_high_en},
respectively. No previous data for this observable exist.
As before, our data are compared to the free (proton+neutron) cross
section from SAID-MA19, BnGa-2019, MAID-2021 multipole analyses and to the calculation on the deuteron.

The overall trend of the data is fairly well reproduced by all models. From the comparison between the free nucleon
predictions and the deuteron results, it can be noted that,  in general, 
nuclear effects
are quite important for all angles over most of the energy range covered by the present measurements.
They also are more relevant at the lowest
  $\theta^{\rm lab}_{\pi^0}$ values, as predicted in Ref.~\cite{itep2} for the unpolarised differential cross section and,     
apart from the first few energy intervals, they lead to a visible decrease in the absolute value of the cross section.  

As for the total $\Delta\sigma$ observable,
there are significant discrepancies
    between these data and the predictions given by the nuclear deuteron model
    in some parts of the measured angular and energy interval.
    As mentioned above, the reason of these differences is not yet understood
    and further theroretical work is needed to solve this problem. 
    
\subsection{Double polarisation $E$ observable for single $\pi^{0}$ on quasi-free proton}

\begin{figure*}[ht]
\centering
\hskip -1cm
\includegraphics[scale=0.7]{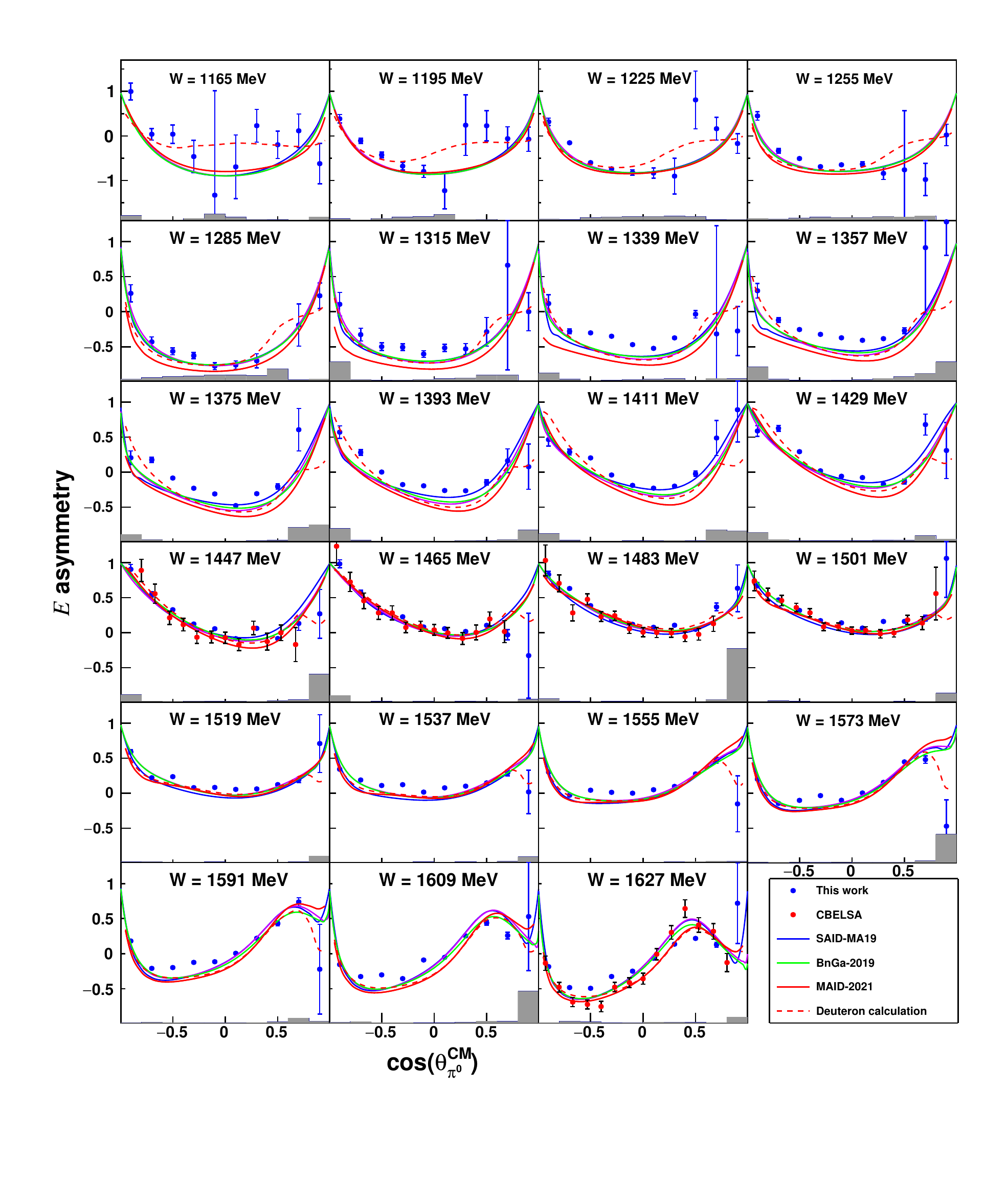}
\vskip -1.75cm
\caption{Helicity asymmetry $E$ for $\pi^0$ photoproduction on quasi-free protons as a function of $\cos(\theta)_{\pi^{0}}^{\mathrm{CM}}$ (blue points) in the energy range $W<1627$ MeV. The pion angle $\theta_{\pi^{0}}^{\mathrm{CM}}$ refers to the $\pi^0p$ center-of-mass frame. The new data are compared to the free proton results reported by CBELSA/TAPS collaboration~\cite{elsa1,elsa2} as well as to the single nucleon calculation with SAID-MA19~\cite{SMA19},
  BnGa-2019~\cite{boga} and MAID-2021~\cite{victor} multipole amplitudes.
  The dashed lines show predictions for quasi-free protons obtained using the model
  of Ref.~\cite{fixar1, fixar2} with the MAID-2021 amplitudes.
The contributions of all
the systematic uncertainties (see Sect.~\ref{par:syserr})  are depicted as grey bars.
}
\label{E_proton_exp_low}
\end{figure*}

\begin{figure*} %
\centering
\hskip -1cm
 \includegraphics[scale=0.7]{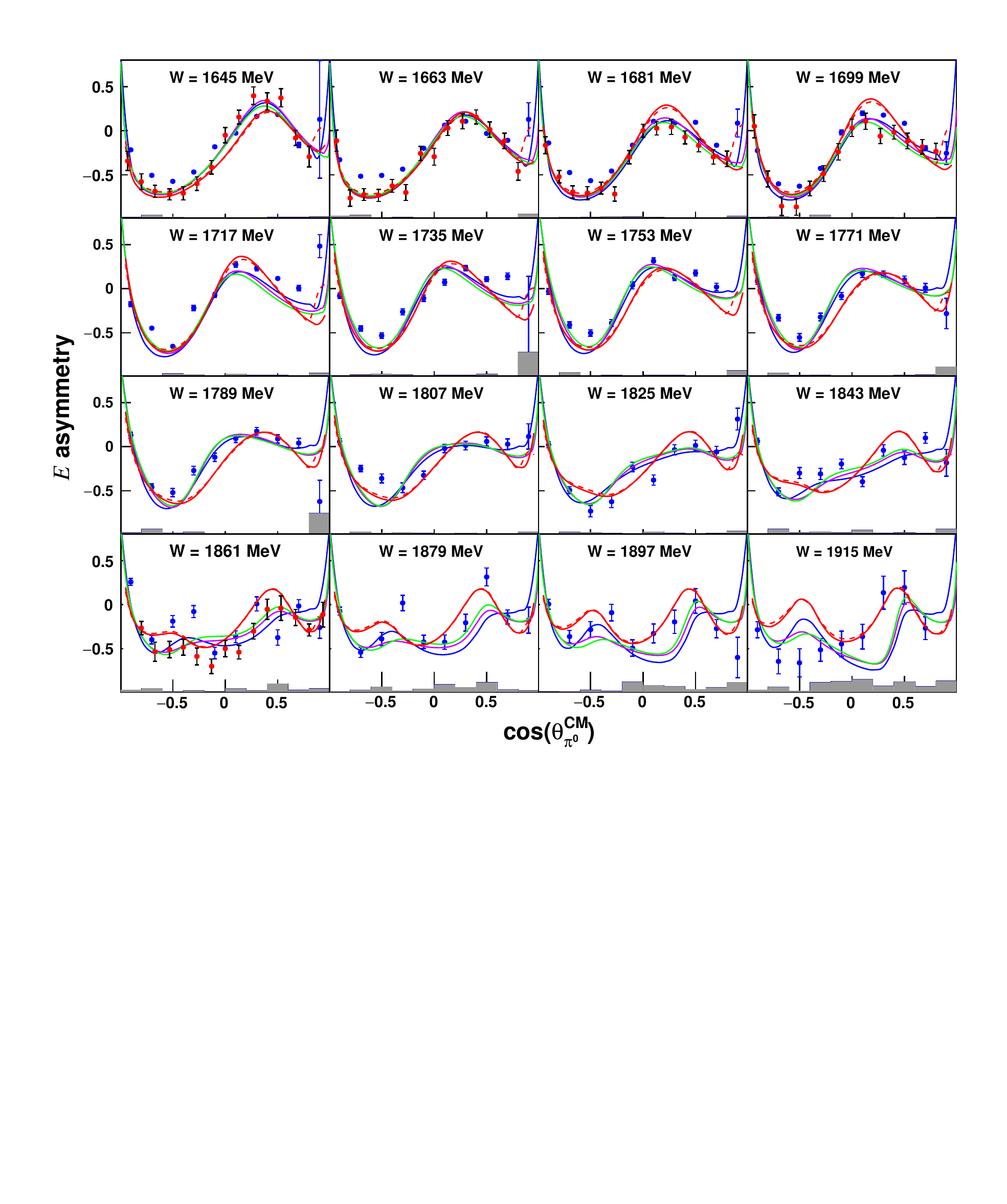}
\vskip -6.5cm
\caption{Same as in Fig.~\ref{E_proton_exp_low} for quasi-free protons and  $W> 1627$ MeV.} 
\label{E_proton_exp_high}
\end{figure*}

The results for the double polarisation observable $E$
on quasi-free protons are presented in Figs.~\ref{E_proton_exp_low} and~\ref{E_proton_exp_high},
where they are compared to the free proton results reported by CBELSA/TAPS collaboration~\cite{elsa1,elsa2}
    when the difference between the measured central $W$ bin values is less than 8 MeV.

The different solid lines represent
 free proton predictions from SAID-MA19 (blue curves),
 BnGa-2019 (green curves) and MAID-2021 (red curves) multipole analyses
     which are constrained by the CBELSA/TAPS data.

The dashed red lines are predictions of the model of Refs.~\cite{fixar1,fixar2}, where  the most important nuclear effects, such as Fermi motion, presence of the $D$-state in the deuteron wave function, Pauli exclusion principle, and first-order rescattering of the final particles, are taken into account.

    The results obtained for the free and the quasi-free proton targets are rather close to each other and agree within statistical and systematic uncertainties. This is an indication
that the nuclear effects have little impact on this observable, at least under the quasi-free kinematic conditions. This fact is by no means trivial considering the strong influence of FSI, shown in Figs.~\ref{Diff_cross_sec_low_en} and  \ref{Diff_cross_sec_high_en}.
Thus, although the cross sections $\sigma^{\uparrow\downarrow/\uparrow\uparrow}$
themselves undergo a noticeable influence from the nuclear environment, this effect tends to almost completely cancel out in the ratio (Eq.~\ref{E_obs2}). This feature is confirmed by the calculations on a deuteron (red dashed lines) which turn out to be very close to the free nucleon results over the major part of the energy range.

Discrepancies occur only in the low $W$ bins and at very forward pion polar angles, for which, as the direct calculation shows, the interaction between the final nucleons is mainly responsible. Namely, for $W$ 
up to about 1300~MeV, 
the detection momentum threshold for nucleons  ($p_{N} \gtrsim 350$~MeV/c) leads to a significant decrease of the phase-space available for quasi-free kinematics. As a result, a substantial fraction of the detected events comes from the kinematical region where the nuclear effects become relevant. 

At the same time, as can be seen from the same figures, despite the rather low statistical accuracy in this $W$ region, the model of Ref.~\cite{fixar2} is able to reproduce the data quite well.

These new results are then particularly important for the neutron case. Since  nuclear effects are basically isospin-independent, one can expect that photoproduction from bound neutrons in
quasi-free kinematics can be used to extract the cross section on a free neutron, without the need to take into account different model-dependent corrections, at least above the first resonance region.


\subsection{Double polarisation $E$ observable for single $\pi^{0}$ on quasi-free neutron}

\begin{figure*} %
\centering
\hskip -1.cm
\includegraphics[scale=0.7]{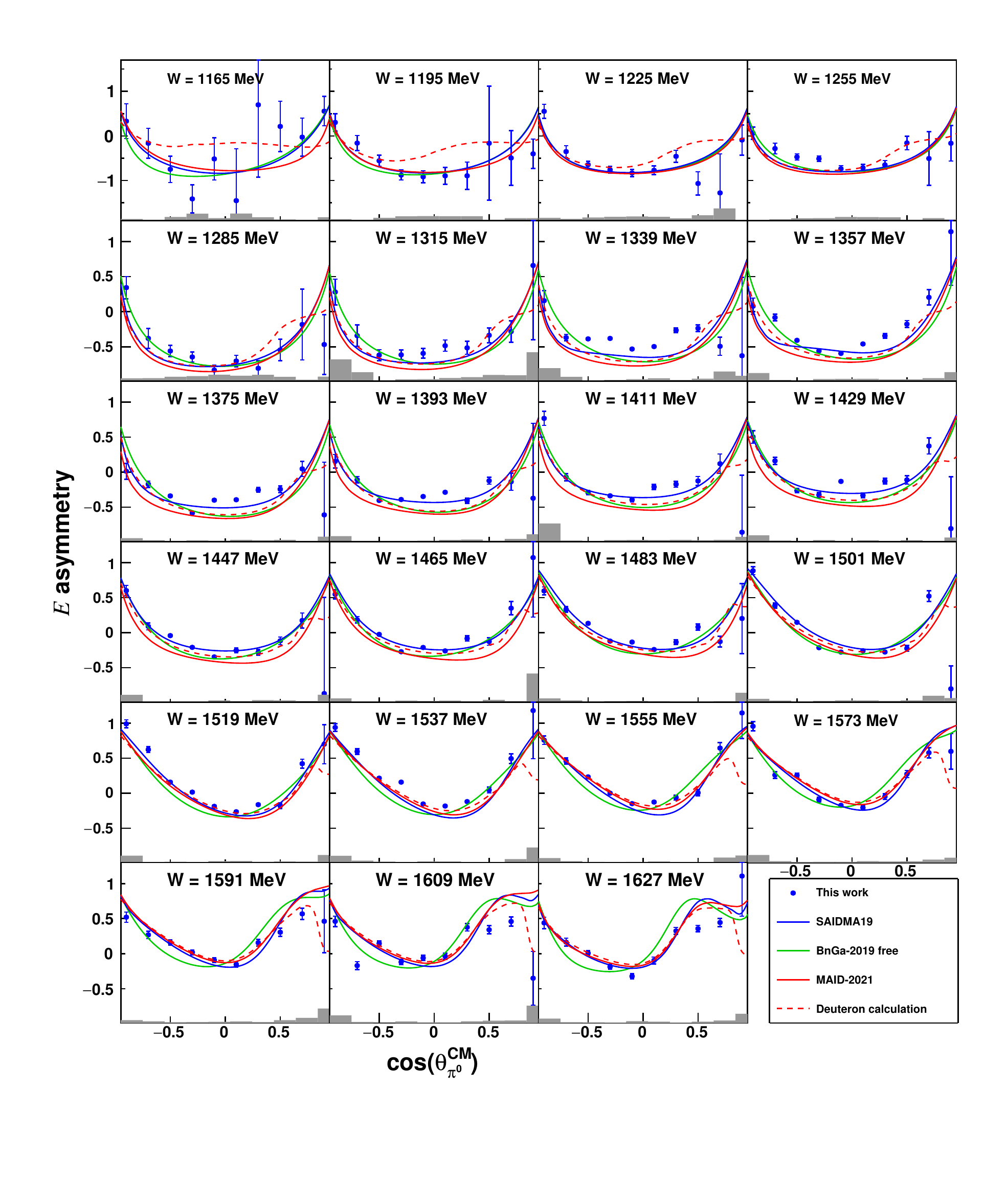}
\vskip -1.75cm
\caption{Same as in Fig.\,\ref{E_proton_exp_low} for quasi-free neutrons.}
\label{E_neutron_1}
\end{figure*}

\begin{figure*} %
  \centering
  \hskip -1.cm
\includegraphics[scale=0.7]{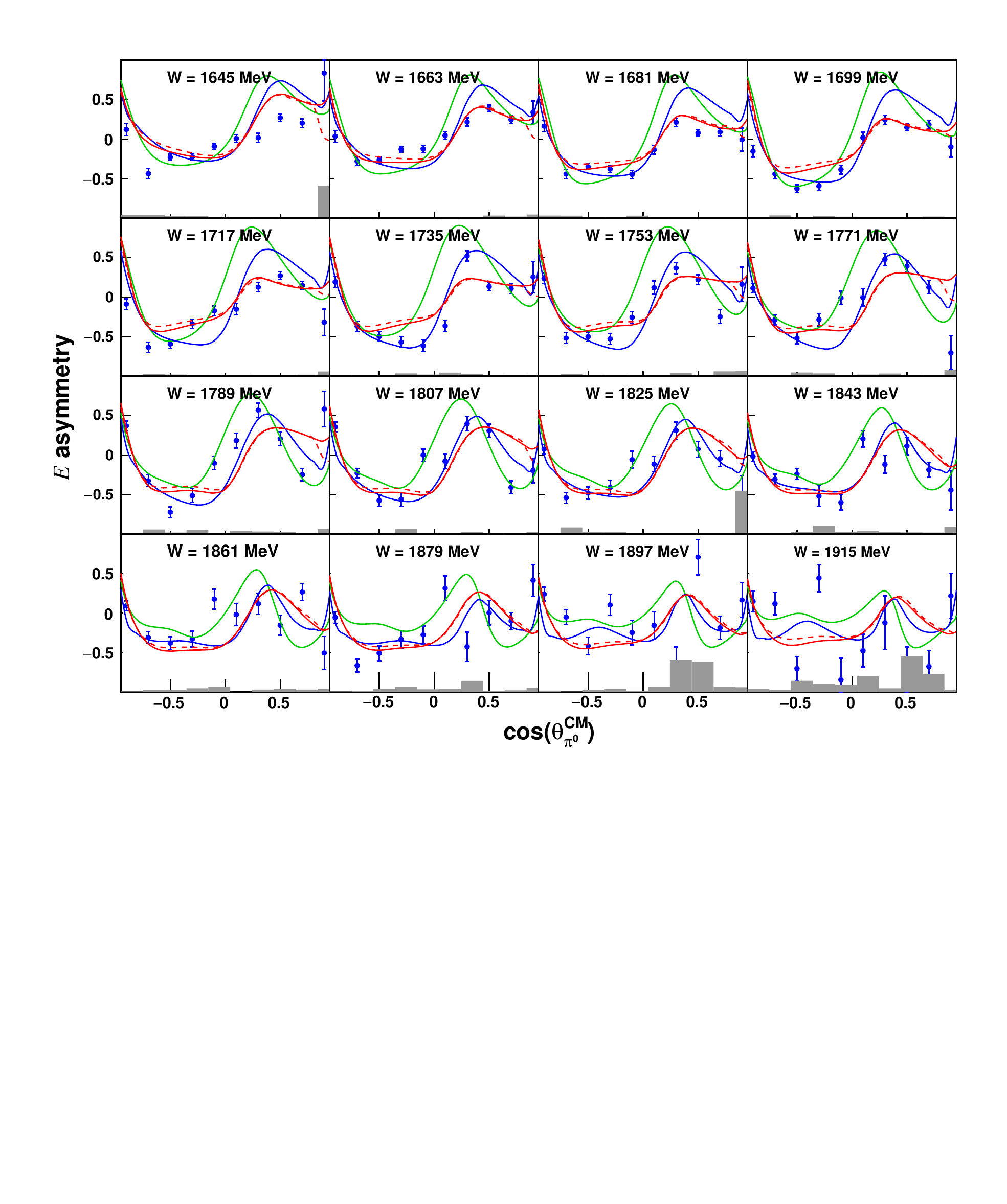}
\vskip -6.5cm
\caption{Same as in Fig.\,\ref{E_neutron_1} for quasi-free neutrons and $W> 1627$ MeV.} 
\label{E_neutron_2}
\end{figure*}

The results of the double polarisation observable $E$ for the single $\pi^{0}$ on quasi-free neutron are shown in Figs.\,\ref{E_neutron_1} and \ref{E_neutron_2},
alongside with the theoretical predictions from the range of models described above. In this case, the present $E$ data have been already included in the data base used to obtain the MAID-2021 predictions. 


These are the first data on the angular distribution of the $E$ observable on  the neutron. The results for the angle-integrated $E$ observable have already been published in Ref.~\cite{diet1}.

The nuclear model predictions show the same features of the proton case. This opens  the possibility of obtaining access to the free-neutron information.
%
%
These new data will then be of great importance to solve the existing discrepancies among the existing multipole analyses that can be noted at $W \gtrsim 1400$~MeV.

\begin{figure*} %
\centering
\includegraphics[scale=0.9]{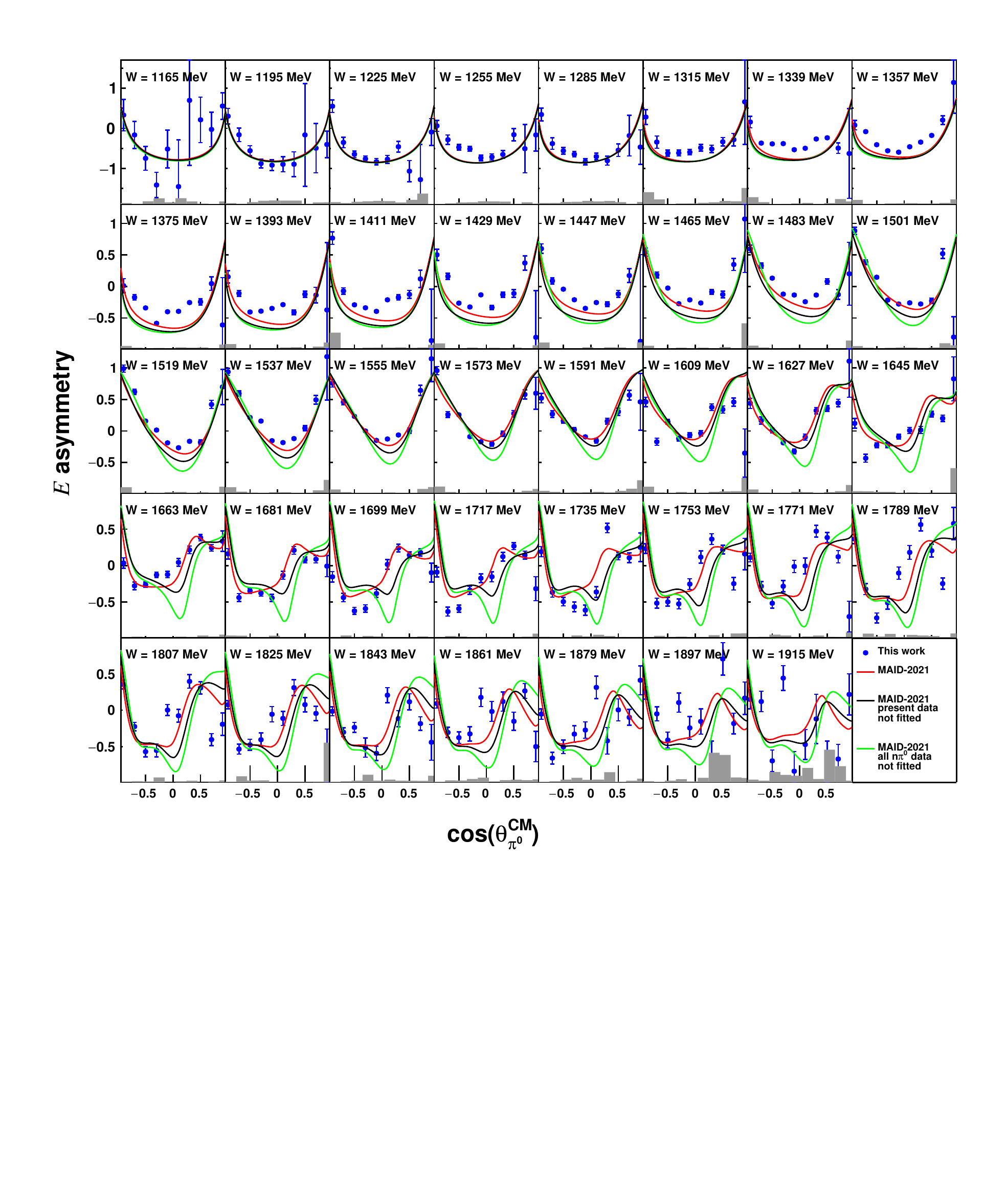}
\vskip -6.0cm
\caption{
MAID-2021 prediction for the helicity asymmetry 
$E$ on the neutron from the fits made with (red line) and  without (black line) including the present data into the fitted data base. The green line is the prediction obtained by excluding from the fit all the published $n\pi^0$ data. }
\label{e-impact}
\end{figure*}

A quantitative evaluation of the impact of these new data on these analyses can be 
obtained with the comparison of the predictions for the $E$ observable on the neutron
from fits made without and  with their inclusion in the full data base.
This comparison is shown in Fig.~\ref{e-impact} using the MAID-2021 analysis. Using our new data 
a relevant change in the predictions can be seen at about $W > 1600$~MeV.

    At lower energies, where different data sets on different observables
    are available,
    the present data, as  could reasonably be expected,
    do not significantly change the predictions given by
    the MAID-2021 partial wave analysis.

\section{Legendre fit of the $E$ data}

To gain a better insight into the partial wave content of the reaction
amplitude,
one can also use expansion of the observables in Legendre polynomials.
Such expansion can be very useful since the energy dependence of the expansion coefficients may reveal specific correlations between individual resonance states of definite parities (see, for instance,Ref.~\cite{wund} and references therein). This method turns out to be especially effective in those cases when a single resonance (for example, $\Delta(1232)$) with well-known properties dominates the amplitude in a certain energy range.

The Legendre coefficients $a_k$ were obtained by fitting the angular distributions of the $E$ asymmetry with a series of associated Legendre polynomials $P_k$:
\begin{eqnarray}
\check{E}(W,\theta)&=&E(W,\theta)\cdot\frac{d\sigma_0}{d\Omega}(W,\theta)=\nonumber \\
&=&{\displaystyle \sum_{k=0}^{2l_{max}}(a_{l_{max}})_k(W)P_{k}(\cos\theta)\,.}
\label{Leg_fit}
\end{eqnarray}
Here, the notation $(a_{lmax})_{k}$ means that in the fitting procedure only the partial waves with the $\pi N$ relative angular momentum up to $l=l_{max}$ were included. The multipoles contributing to the fit for $l_{max}=1,2,3$ are listed in Table \ref{tab:Sensibility-on-multipoles}.  

\begin{table}[ht]
\centering{}
\captionsetup{width=0.95\linewidth}
\caption{The multipole amplitudes contributing to the fitted cross section
  reported in Eq.~(\ref{Leg_fit})
  for different choices of $l_{max}$.  \label{tab:Sensibility-on-multipoles}}
\begin{tabular}{|c|c|c|}
\hline 
$l_{max}$ & wave & M-poles\tabularnewline
\hline 
\hline 
\multirow{2 }{* }{1} & S-wave & $E_{0^+}$
\tabularnewline
 & P-wave & $E_{1^+},M_{1^+},M_{1^-}$\tabularnewline
\hline 
2 & D-wave & $E_{2^+},E_{2^-},M_{2^+},M_{2^-}$\tabularnewline
\hline
3 & F-wave & $E_{3^+},E_{3^-},M_{3^+},M_{3^-}$\tabularnewline
\hline 
\end{tabular}
\end{table}

For the unpolarised cross section $d\sigma_0/d\Omega$ in Eq.\,(\ref{Leg_fit}), we used the values given by the SAID-MA19 analysis. The latter are in good agreement with the available unpolarised data both on the proton and on the neutron.
Replacing the SAID-MA19 analysis with the BnGa-2019 analysis gives almost the same results for $a_k$ within statistical uncertainties.

%

\begin{figure*} %
\centering
\includegraphics[scale=0.65]{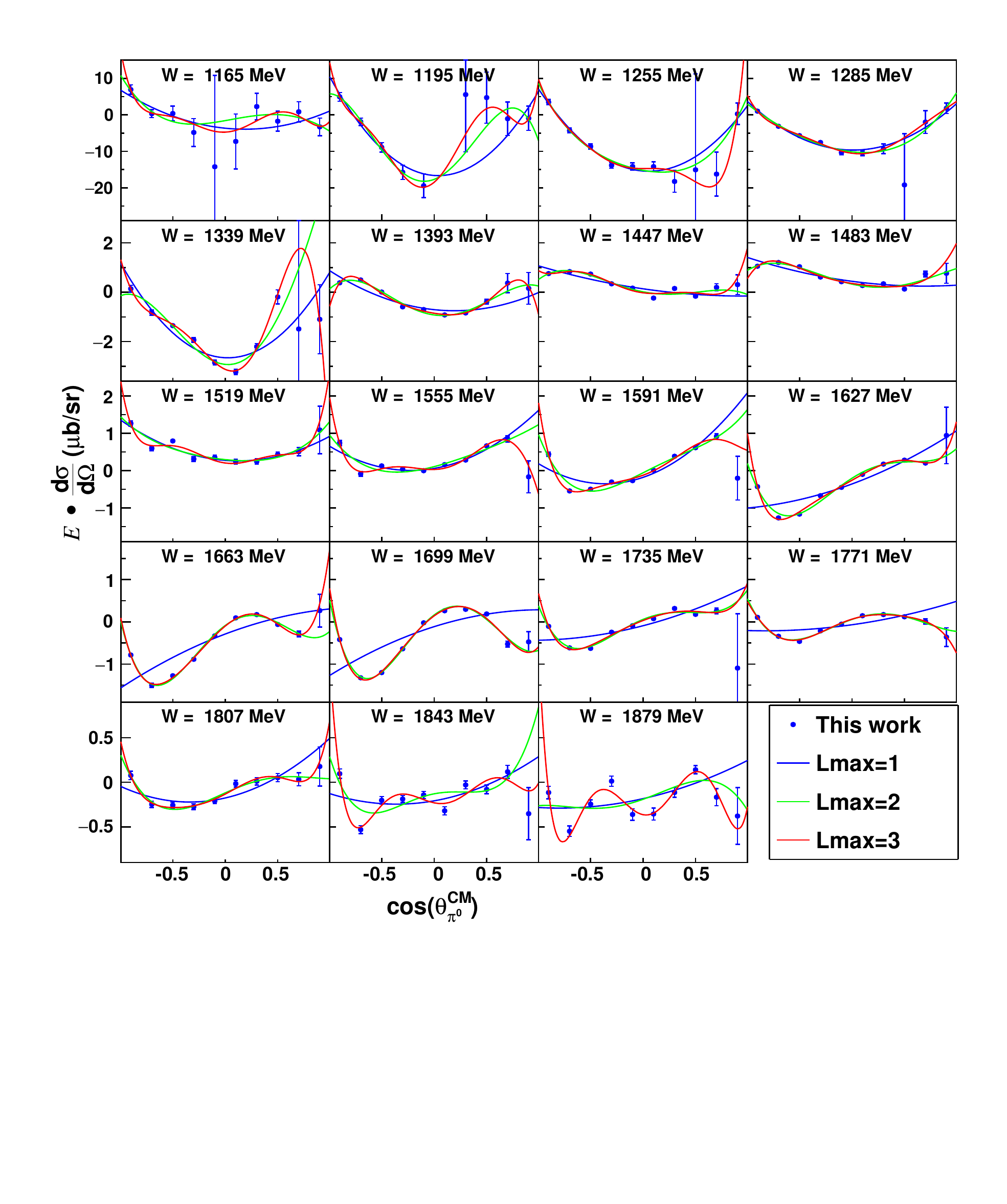}
\vskip -3.25 cm
\caption{The asymmetry $\check{E}=E\,d\sigma_0/d\Omega$ as function of $\cos(\theta)_{\pi^{0}}^{\mathrm{CM}}$
  in the  $\pi^{0}p$ channel. Different fits are obtained with Legendre polynomial
  expansion given in Eq.~(\ref{Leg_fit}) 
truncated at $l_{max}=$ 1 (blue), 2 (green), and 3 (red).}
\label{fit_proton_legendre}
\end{figure*}

\begin{figure*} 
\centering
\includegraphics[scale=0.65]{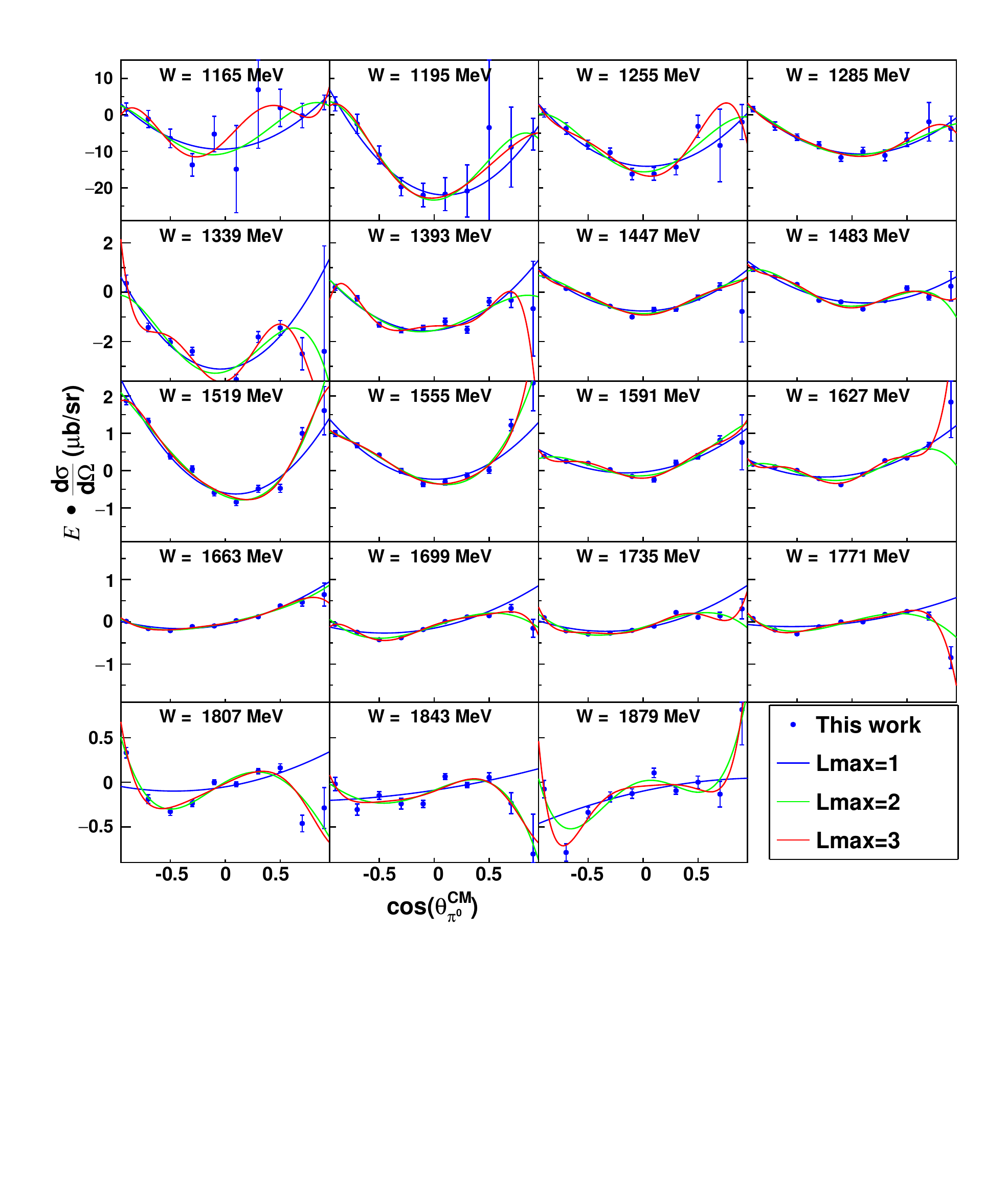}
\vskip -3.25 cm
\caption{Same as in Fig.\,\ref{fit_proton_legendre} for the $\pi^{0}n$ channel.}
\label{fit_neutron_legendre}
\end{figure*}

The quality of our fit with $l_{max}=1,2,3$ is demonstrated in Figs.\ \ref{fit_proton_legendre} and \ref{fit_neutron_legendre} for several values of $W$.
In the region $W<1400$\,MeV, where the $\Delta(1232)$ resonance dominates, the angular dependence of $\check{E}$  should be governed by the $p$-waves with relatively small admixture of the $s$-waves. The smallness of the $s$-wave part is explained by the relative weakness of the electric dipole amplitude $E_{0+}$, which is responsible for production of the $s$-wave pions. In the $\pi^0$ channel,
the $s$-wave
is an order of magnitude smaller than the charged pion one. As a result, in the wide energy range up to the second resonance region, the $\pi^0$ photoproduction proceeds almost exclusively via the magnetic dipole transition to the $\Delta(1232)$ resonance. There is also a small admixture from the nucleon pole terms in the direct and crossed channels from the magnetic $\gamma N$ coupling. Thus, taking $l_{max}=1$ is expected to be sufficient to describe the  general behavior of the data in a rather wide energy range. This explanation is fairly well supported by the experimental results in Figs.~\ref{fit_proton_legendre} and \ref{fit_neutron_legendre}.

Above $W=1400$\,MeV, the second resonance, $N(1520)$, starts to play a role, so that $l$ needs to be expanded to $l_{max}=2$ in order to take into account an increasing contribution of the $D$-waves.


Since the $\chi^2$ value does not change significantly  when going from $l_{max}=2$  to $l_{max}=3$ and only ten data points are available, $l_{max}=2$ has been chosen as the best compromise between fit efficacy and our partial-wave analysis.

The investigation of the expansion given in Eq.~(\ref{Leg_fit})) reveals some important properties of the Legendre coefficients $a_k$. Firstly, parity conservation requires that the coefficients $a_k$ with even~$k$ contain the products of multipoles $A_{l^\pm}A^\prime_{l^{\prime\pm}}$ (with $A_{l^\pm}=E/M_{l^\pm}$) for which the difference ($l-l^\prime$) takes only even values. Accordingly, the coefficients with odd~$k$ include products in which this difference is odd. This means that the odd coefficients are determined exclusively by the interference of the resonances with different parities. This property explains, in particular, the relative smallness of these coefficients over the entire energy range. 

Another important property of the expansion coefficients is that the terms of the type $|E/M_{l^-}|^2$, quadratic in multipoles with total spin $j=l-1/2$, contribute only to the coefficients $a_0$, ..., $a_{2l-2}$ and do not appear in $a_{2l}$. This is a consequence of the total angular momentum conservation. For this reason, for example, the coefficient $a_4$ does not contain the terms 
$|E_{2^-}|^2$, $|M_{2^-}|^2$, and $E^*_{2^-}M_{2 -}$ of the resonance multipoles coming from $D_{13}(1520)$.

\begin{figure*} 
\centering
\includegraphics[scale=0.75]{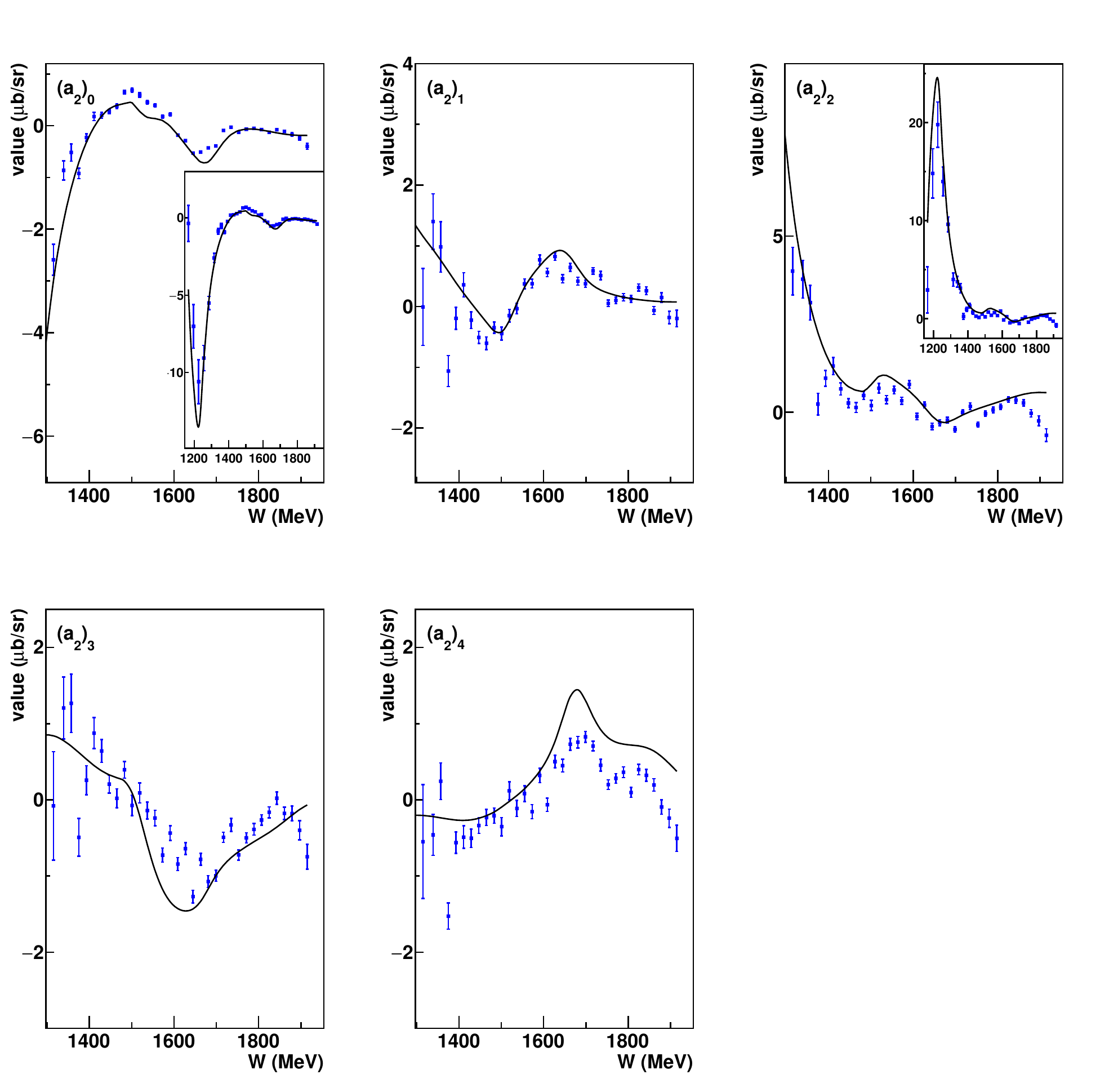}
\caption{Comparison of the fitted Legendre coefficients $(a_2)_k$ for the $\pi^0p$ channel 
with the SAID-MA19 model predictions.}
\label{fit_5parameter_proton}
\end{figure*}

\begin{figure*} 
\centering
\includegraphics[scale=0.75]{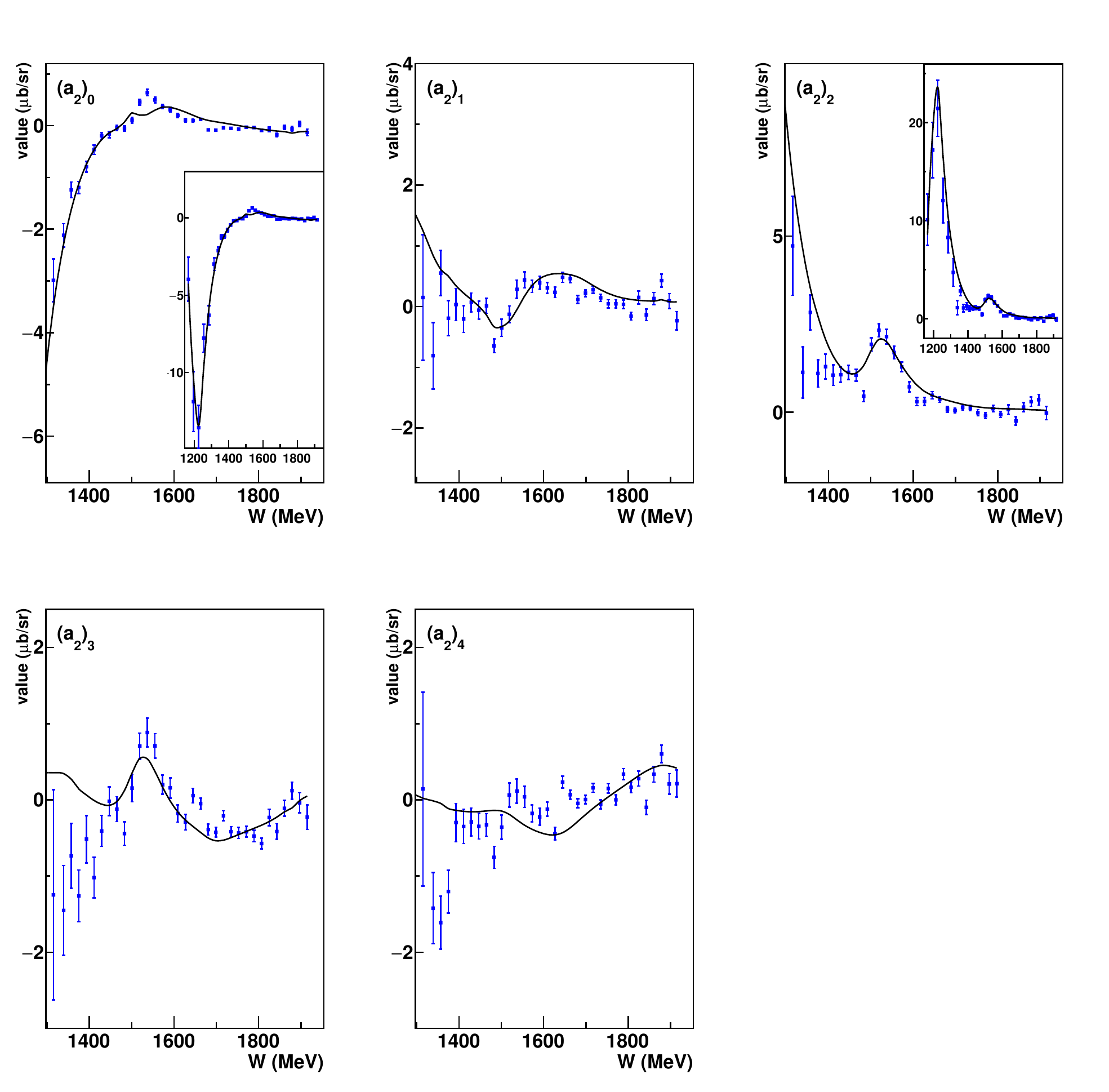}
\caption{Same as in Fig.\,\ref{fit_5parameter_proton} for the $\pi^0n$ channel.}
\label{fit_5parameter_neutron}
\end{figure*}

%
Following from the discussion above, up to the energy $W \simeq 1400$\,MeV the reaction is dominated by the $p$-wave multipole $M_{1^+}$ due to the $\Delta(1232)$ excitation mechanism. A direct consequence of this dominance is a pronounced resonance to be expected in the coefficients $a_0$ and $a_2$ in the energy region around $W=1230$\,MeV,
with all the remaining coefficients having very small values.
This expected resonance behavior is observed. See inserts in $(a_2)_0$ and $(a_2)_2$ plots
in Figs.~\ref{fit_5parameter_proton} and \ref{fit_5parameter_neutron}.

%

The Legendre coefficients for the proton and the neutron channels given by  the
 fit procedure described above 
are plotted in Fig.~\ref{fit_5parameter_proton} and in Fig.~\ref{fit_5parameter_neutron}, respectively, for 
$W > 1300$~MeV, a region where, as discussed above, nuclear effects are minimized and all the coefficient values are significantly different from zero. %

In the insert plots of the fitted $(a_2)_0$ and $(a_2)_2$ coefficients, the only ones that have meaningful values in the $\Delta (1232)$ resonance region, their values are given over the full measured $W$ range.
The curves represent the corresponding coefficients evaluated using the SAID-MA19 model.

It is interesting to note that the coefficient $(a_2)_0$  exhibits, as expected, a cusp structure at the $\eta$ threshold 
($W=$ 1487~MeV) in both the  $\pi^0p$ and the $\pi^0n$ channels.
    This effect, previously observed  for the $\Delta\sigma$ observable
    (see Fig.~\ref{total_pol_cs}), is 
due to interference of the negative parity state $S_{11}(1535)$ with the $\Delta(1232)$ resonance. 
For the same coefficient, the effect of the intermediate excitation of the $F_{15}(1680)$ resonance is clearly visible at higher energies in the proton case.
This could be predicted from the much smaller absolute value of the $A_{3/2}$ helicity amplitude of this resonance in the neutron case.
%

As discussed above, the $(a_2)_4$ coefficient does not contain the terms $A^*_{2^-}A^\prime_{2^-}$, $A=E/M$, that are determined by the $D_{13}$ wave alone. In the case $l_{max}=2$ its value is due only to the interference of the $E/M_{2^-}$ and the $E/M_{2^+}$ multipoles. The almost complete absence of the resonance-like structure 
around $W=1500$~MeV is a trivial consequence of the smallness of the $E/M_{2^+}$ amplitudes in this energy region.

The structure in the data at $W>$ 1600~MeV, which is especially evident for the proton case, is due to the fact that our fit procedure is limited to $l_{max}=2$. This  artificially increases the contribution of the $D_{13}(1520)$ and  $D_{15}(1650)$ to compensate the real effect due to the onset of $F_{15}(1680)$ resonance. As mentioned before, due to the limited number of angular bins and also to the limited statistical accuracy in the polar forward region, this contribution can not be properly evaluated. 

In the future,
new experiments with higher statistics are required,
and a more careful evaluation of nuclear effect is needed, in particular
at very forward polar angles, as suggested  by  calculation of Refs.~\cite{fixar1,fixar2}.

%
%
%


  %


\section{Summary and conclusions}
New precise data on the helicity-dependent inclusive cross section, as well as on the beam-target helicity spin asymmetry $E$ of single $\pi^0$ photoproduction on the de\-u\-te\-ron, have been obtained. 
Compared to existing data, the new measurements cover a wider energy range and have
higher statistical precision.

Comparison with the free nucleon calculation allows 
the influence of nuclear effects on the single photoproduction mechanism to be evaluated quantitatively.
This in turn gives valuable information about the extent to which these effects may distort the helicity asymmetry $E$, extracted from the quasi-free nucleon cross sections.

According to our results, the difference $d\Delta\sigma/d\Omega$
(see Eq.~\ref{eq:beam-recoil})
of the helicity-dependent cross sections $\sigma^{\uparrow\downarrow/\uparrow\uparrow}$ exhibits
rather different behaviors between free nucleons  and nucleons inside deuterium.
%
These differences are not correctly reproduced by the 
    nuclear deuteron model for some parts the measured angular and energy range
    and further theoretical work is needed to provide
    a better understanding of the nuclear effects and to link nucleon
  properties  to nuclear 
properties.

At the same time, in the asymmetry $E$ values measured for quasi-free nucleons, the nuclear effects are to a relevant extent canceled and can be disregarded in most
of the measured energy and angular range.
Exceptions are
for  energies below the $\Delta(1232)$ peak and, as also observed in Ref.~\cite{mullen} for the photon beam asymmetry,
at very forward pion angles.

Therefore, these new data on $E$ for quasi-free  neutrons can be used to access this observable on the free neutrons without resorting to any model-dependent calculations.

    Our new results will  have a important effect in resolving the discrepancies
    between the existing multipole analyses and a better understanding of the neutron excitation, especially  for $W \gtrsim 1600$~MeV, where discrepancies are more prounounced and, as seen for MAID-2021, their impact is more significant. 

A Legendre analysis of the new experimental results has already provided,
without performing a detailed partial wave analysis,
valuable information on the 
resonance states contributing to the $\pi^0n$ channel, where the role 
played by the $F_{15} (1680)$ resonance turned out to be smaller than in the
proton case. 

\section{Acknowledgments}

The authors wish to acknowledge the excellent support of the accelerator group of MAMI.
 This work has been supported by the U.K. STFC (ST/L00478X/1, ST/T002077/1, ST/L005824/1, 57071/1, 50727/1,\\ ST/V001035/1) grants, the Deutsche Forschungsgemeinschaft (SFB443, SFB/TR16, and SFB1044), DFG-RFBR (Grant No. 09-02-91330), Schweizerischer Nationalfonds (Contracts No. 200020-175807, No. 200020-156983, No. 132799, No. 121781, No. 117601), the U.S. Department of Energy (Offices of Science and Nuclear Physics, Awards No. DE-SC0014323, DEFG02-99-ER41110, No. DE-FG02-88ER40415, No. DEFG02-01-ER41194) and National Science Foundation (Grants NSF OISE-1358175; PHY-1039130, PHY-1714833, No. IIA-1358175), INFN (Italy), and NSERC of Canada (Grant No. FRN-SAPPJ2015-00023). 
 
 \bibliographystyle{h-elsevier}

%

\end{document}